\documentclass[useAMS,usenatbib]{mn2e_mod}
\usepackage{deluxetable}
\usepackage[utf8]{inputenc}
\usepackage{amsmath}
\usepackage{amsfonts}
\usepackage{amssymb}
\usepackage{graphicx}
\usepackage[font=small,labelfont=bf]{caption}
\usepackage[a4paper]{geometry}
\usepackage{tablefootnote}
\usepackage{hyperref}
\usepackage{natbib}
\title {A Sample of Type II-L Supernovae}

		\author[Faran et al.]
		{T. Faran$^{1}$, 
		D. Poznanski$^{1}$\thanks{dovi@tau.ac.il},
		A. V. Filippenko$^{2}$,
		R. Chornock$^{3}$,
		R. J. Foley$^{4}$,\newauthor
		M. Ganeshalingam$^{2,5}$,		
		D. C. Leonard$^{6}$,
		W. Li$^{2,7}$,
		M. Modjaz$^{8}$, 
		F. J. D. Serduke$^{2}$,\newauthor
		and J. M. Silverman$^{9,10}$\\
		\\
		$^{1}$School of Physics and Astronomy, Tel-Aviv University, Tel Aviv 69978, Israel.\\
		$^{2}$Department of Astronomy, University of California, Berkeley, CA 94720-3411, USA.\\
		$^{3}$Harvard-Smithsonian Center for Astrophysics, 60 Garden Street, Cambridge, MA 02138, USA.\\
		$^{4}$Department of Astronomy, University of Illinois, Urbana, Illinois 61801, USA.\\
		$^{5}$Lawrence Berkeley National Laboratory, Berkeley, CA 94720, USA.\\
		$^{6}$Department of Astronomy, San Diego State University, San Diego, CA 92182, USA.\\
                $^{7}$Deceased 12 December 2011.\\
		$^{8}$CCPP, New York University, 4 Washington Place New York City, NY, 10003, USA.\\
		$^{9}$Department of Astronomy, University of Texas at Austin, Austin, TX 78712, USA.\\
		$^{10}$NSF Astronomy and Astrophysics Postdoctoral Fellow\\
		}

\begin{document}
	\maketitle
	\label{firstpage}
	\begin{abstract}
What are Type II-Linear supernovae (SNe~II-L)? This class, which has been ill defined for decades, now receives significant attention --- both theoretically, in order to understand what happens to stars in the $\sim 15$--25\,M$_{\odot}$ range, and observationally, with two independent studies suggesting that they cannot be cleanly separated photometrically from the regular hydrogen-rich SNe~II-P characterised by a marked plateau in their light curve. Here, we analyze the multi-band light curves and extensive spectroscopic coverage of a sample of 35 SNe~II and find that 11 of them could be SNe~II-L. The spectra of these SNe are hydrogen deficient, typically have shallow H$\alpha$ absorption, may show indirect signs of  helium via strong O\,I $\lambda$7774 absorption, and have faster line velocities consistent with a thin hydrogen shell. The light curves can be mostly differentiated from those of the regular, hydrogen-rich SNe~II-P by their steeper decline rates and higher luminosity, and we propose as a defining photometric characteristic the decline in the $V$ band: SNe~II-L seem to decline by more than 0.5 mag from peak brightness by day 50 after explosion. Using our sample we provide template light curves for SNe II-L and II-P in 4 photometric bands. 

\end{abstract}
\begin{keywords}
Supernovae: general
\end{keywords}

\section{Introduction}

\begin{figure*}
\centering
\includegraphics[width=1\textwidth]{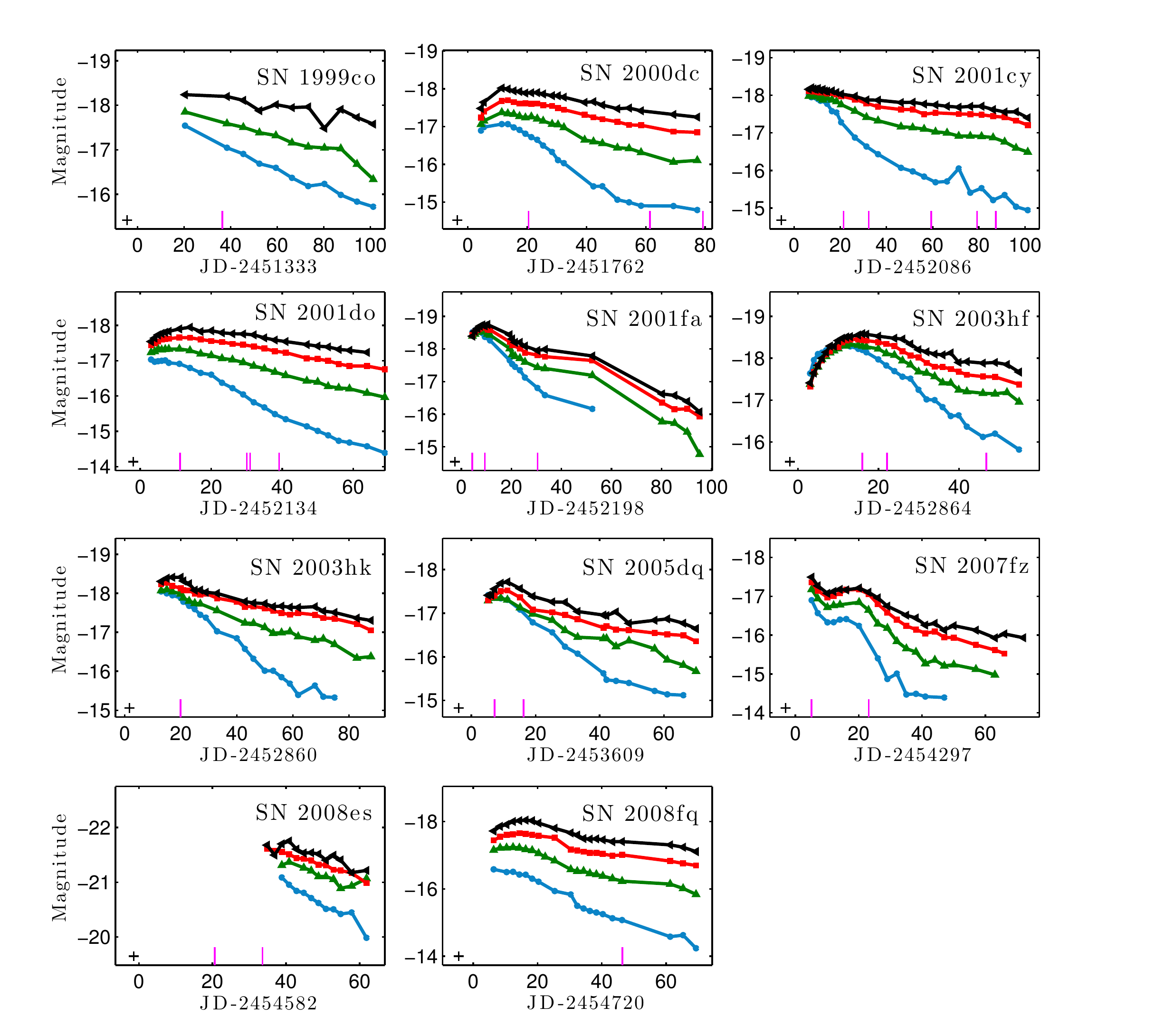}
\caption{Light curves for the 11 SNe in our sample: $B$ (blue circles), $V$ (green triangles), $R$ (red squares), and $I$ (black side triangles). Magenta ticks mark the epochs at which spectra were obtained. The cross signs mark the last nondetection day.}\label{f:phot_grid}
\end{figure*}

Core-collapse supernovae (SNe) are divided into an array of subtypes based on their observational properties. This could be seen as a trivial and pedantic exercise; however, when ultimately trying to decipher the fate of massive stars, as well as their interplay with their surroundings, one needs to be able to draw a clear mapping from progenitor to SN. The accepted classification scheme is mostly spectroscopic, based on the presence of hydrogen (Type II) or lack of it (Types Ib and Ic). Some Type II SNe are further assigned a subtype based on the presence of relatively narrow emission lines (IIn) or having intermediate properties --- starting off as a Type II, evolving into a Ib (IIb). To further complicate the picture, there is a photometric distinction between SNe~II-P that have a long period of roughly constant luminosity (a plateau) and SNe~II-L with a linearly (in magnitudes) declining light curve (see \citealt{Filippenko:1997} for a review). 

Clearly, spectroscopic classes and photometric classes are not necessarily mutually exclusive. An object could in principle have a linearly declining light curve, supposedly making it a SN~II-L, and yet spectroscopically resemble other classes (IIb or IIn). The spectroscopic signature of SNe~II-L (if indeed they form an independent class) is ill defined. [An exception to this is tentative evidence that SN~II-L spectra do not show a well-developed H$\alpha$ absorption component in the P-Cygni line profile \citep{Schlegel:1996}.] Thus, mostly owing to the time required to acquire, calibrate, and analyze a light curve (as opposed to the nearly instantaneous spectroscopic analysis), the SN~II-L class has been largely unassigned for decades.

A major impediment to detailed studies of SNe~II-L is that they are rare, though not as rare as might appear from a casual literature search. Out of 1154 unambiguous Type II classifications reported to the IAU between the years 2000 and 2013, 697 are not assigned a subtype, 228 are II-P, 158 are IIn, 68 are IIb, and only 3 are II-L. This is a direct consequence of the lack of a spectroscopic definition of SNe~II-L, and it suggests a rate among SNe~II of less than a percent. However, having redefined SNe~II-L as objects that decline by $\sim 1$ mag over 100 days \citep{Poznanski:2009}, \citet{Li:2011} find that while SNe~II-P comprise about 60\% of all core-collapse events, SNe~II-L still constitute a few percent (see also \citealt{Smartt:2009}). 

The prevalence of SNe~II-P is compatible with our current understanding that their progenitors are also the most common massive stars (the least massive ones): red supergiants (RSGs) with masses of 8--15\,M$_{\odot}$, a link strongly established by progenitor detections \citep{Smartt:2009}. In contrast, there is only weak evidence regarding the progenitors of SNe~II-L. SN\,1979C (e.g., \citealt{Branch:1981}) and SN\,1980K (e.g., \citealt{Uomoto:1986}) are the nearest and best-observed ``historical'' SNe~II-L. The position of SN\,1980K was imaged 49 days before maximum light \citep{Thompson:1982}, but no star was found to be coincident with the position of the SN,  ruling out RSGs more massive than 20\,M$_{\odot}$ though allowing bluer progenitors with $T >$ 10,000\,K and a mass in the range of 10--15\,M$_{\odot}$ \citep{Smartt:2009}. An analysis of the stellar population around SN\,1979C shows that with the exception of one supergiant, most of the stars appear to be massive main-sequence turnoff stars. Hence, the progenitor of SN\,1979C was likely a RSG with mass $\sim 17$--18\,M$_{\odot}$ \citep{vandyk:1999}. 

The only ``modern'' (i.e., post \emph{Hubble Space Telescope}) SN~II-L with a progenitor detected is SN\,2009kr (\citealt{Elias-Rosa:2010}; \citealt{Fraser:2010}). There is debate regarding the interpretation of the observations. While \citet{Elias-Rosa:2010} favor a massive progenitor of 18--24\,M$_{\odot}$, \citet{Fraser:2010} argue that if the progenitor is a single star, then it is a yellow supergiant with a mass of $15^{+5}_{-4}\,{\rm M}_{\odot}$. Based on the mass-velocity relation  (which suggests that the progenitor mass is proportional to the day 50 photospheric velocity, $M \propto v$), \citet{Poznanski:2013} argue that the higher mass is favored. 

\vspace{1cm}

The distribution of SN~II-P progenitor masses (see the compilation by \citealt{Smartt:2009} and Table~1 of \citealt{Poznanski:2013}) spans masses approximately between 8\,M$_{\odot}$ and 18\,M$_{\odot}$ with an upper limit of 25\,M$_{\odot}$. The few SN~II-L progenitor masses are thus at the higher end of this distribution, perhaps just beyond it.

There is a paucity of progenitor detections with masses greater than $\sim 15$\,M$_{\odot}$, as well as an observed gap in the the masses of stellar remnants. Neutron star (NS) masses reach about 2M$_{\odot}$, while black hole (BH) masses seem to start near 5\,M$_{\odot}$ \citep{Ozel:2010,Ozel:2012}. A compelling solution to these two mysteries was suggested (see \citealt{Kochanek:2013}, and references therein). If above some mass threshold stars fail to produce a SN, and collapse directly to form a BH, they accumulate more mass in the core, thus explaining the perceived lack of progenitors as well as the gap in remnant masses. Based on the rate determinations of \citet{Li:2011}, \citet{Smith:2011} argue that there is no gap when translating the SN rates to the initial mass function; however, \citet{Kochanek:2013} dismisses this argument, claiming that the uncertainty in the fraction of SNe coming from binaries could allow for the needed adjustment. This discussion is significantly hampered by the uncertain definition and poorly studied properties of SNe~II-L. Understanding SNe~II-L, their progenitors, and where they fit in the progenitor-to-SN mapping is our main purpose here.\footnote{This paper is dedicated to the memory of our dear friend and colleague, Dr. Weidong Li, without whom these data would not exist.}. 

As one goes up the mass function, wind-induced mass loss becomes a dominant process, particularly during the RSG phase. In fact, models show that stars more massive than $\sim 20$\,M$_{\odot}$ can end up with less mass than their less-massive siblings, after having lost most or all of their hydrogen outer shells. The remaining thin hydrogen envelopes are further helium enriched by convective dredge-up from the helium shell to the outer envelope \citep[e.g.,][]{Kasen:2009}. This has a direct effect on the helium mass fraction, since the helium-core mass is an increasing function of the initial mass. 

The photometric plateau of SNe~II-P is interpreted as being powered by a hydrogen recombination wave propagating inward through the SN envelope, slowly releasing all the energy deposited by the shock (e.g., \citealt{Popov:1993}; \citealt{Kasen:2009}). A thinner hydrogen envelope, as one would expect from a more massive progenitor, cannot support a long plateau phase, releasing the trapped energy faster. Furthermore, the abundance of helium in the envelope alters the electron-scattering opacity and the recombination temperature of the ejecta, since the recombination temperature of helium is higher than that of hydrogen ($\sim$ 10,000--12,000\,K; \citealt{Arnett:1996}). 

In a recent numerical study, \citet{Bayless:2014} tested the effects of mass loss on the light curve of a SN with a progenitor of 23\,M$_{\odot}$ (initial mass). They found that removing hydrogen from the outer shell gradually alters the nature of the light curve from II-P to IIb/Ib, where the decline rate of the light curve correlates with mass loss.

Therefore, SNe from progenitors near $\sim 20$\,M$_{\odot}$ should be more luminous than SNe~II-P, have more rapidly declining light curves, and display evidence of helium, all due to the thin hydrogen shell. As already suggested by \citet{Patat:1994} and as we show below, SNe~II-L consistently fit this picture.

Recently, as this paper was in its final editing stages, two groups published large samples of SNe~II \citep{Anderson:2014,Sanders:2014}. Both found that contrary to previous claims (the latest by \citealt{Arcavi:2012}), one cannot photometrically separate SNe~II-L from SNe~II-P. While our sample is smaller and we cannot make definitive conclusions, we address this issue as well. \citet{Gutierrez:2014} focus on a spectroscopic study of the H$\alpha$ line, using a somewhat larger sample than ours, and find results which are consistent with those we present below.

\section{Observations}

The sample of SNe was compiled from data collected with the 0.76-m Katzman Automatic Imaging Telescope (KAIT), as a part of the Lick Observatory SN search. A sample of 35 SNe~II was constructed, excluding SNe~IIb and IIn based on spectroscopic identifications. In a companion paper \citep[][hereafter F14a]{Faran:2014} we analyze in detail the 23 SNe~II-P in the sample and discuss all of the observational and reduction details. SN\,2000cb, a SN\,1987A analog, is analyzed by \citet{Kleiser:2011}. The remaining 11 SNe are discussed here, and listed in Table \ref{t:SNedataP2}. As explained in detail by F14a, the photometry is corrected for Galactic extinction using the maps of \citet{Schlegel:1998}, but we do not correct for host-galaxy extinction. All photometric and spectroscopic data are available electronically via the Berkeley\footnote{http://hercules.berkeley.edu/database/} \citep{Silverman:2012}  and WISeREP\footnote{http://www.weizmann.ac.il/astrophysics/wiserep/} \citep{Yaron:2012} databases. Table \ref{t:phot_short} includes a sample of the photometric data. The full table is available in the online version. The journal of spectroscopic observations is included in Table \ref{t:spec}. 

\begin{table}
	\tiny
	\tabcolsep=0.11cm
\caption{SN II-L Sample}\label{t:SNedataP2}
\begin{tabular}{lcccc}
\hline\hline
SN name &
$z_{\rm host}$ &
$\mu$ (mag)\tablenotemark{a} &
Explosion MJD &
$E(B-V)_{\rm MW}$\tablenotemark{c}\\
\hline

 1999co  & 0.0314 & 35.52$^{b}$  & 51332 (5) & 0.0490\\
 2000dc  & 0.0104 & 33.52  & 51762 (4) & 0.0796\\
 2001cy & 0.0149 & 33.94$^{b}$  & 52085 (6) & 0.1983\\
 2001do & 0.0104 & 33.16$^{b}$  & 52133 (2) & 0.1950 \\
 2001fa  & 0.0173 & 34.58  & 52198 (3) & 0.0770\\
 2003hf & 0.0313 & 35.55$^{b}$  & 52863 (2) & 0.0220\\
 2003hk  & 0.0226 & 35.07 & 52860 (?) & 0.0335\\
 2005dq  & 0.0219 & 34.77$^{b}$  & 53608 (4) & 0.0803\\
 2007fz  & 0.0144 & 33.86$^{b}$  & 54296 (3) & 0.0402\\
 2008es  & 0.2130 & 39.38$^{b}$  & 54581 (1) & 0.0117\\
 2008fq  & 0.0106 & 32.86  & 54720 (5) & 0.0637\\
\hline
\end{tabular}
\tablenotetext{a}{Distance modulus from NED, unless noted otherwise.}
\tablenotetext{b}{Redshift-based distance.}
\tablenotetext{c}{Assume $R_V=3.1$.}
\end{table}

\begin{table*}
	\tiny
	\tabcolsep=0.11cm
\caption{Sample light curves.\label{t:phot}}
\begin{tabular}{ccccccc}
\hline\hline
SN name &
MJD &
Age \tablenotemark{a} &
\emph{B}($\sigma_B$) &
\emph{V}($\sigma_V$) &
\emph{R}($\sigma_R$) &
\emph{I}($\sigma_I$)\\
\hline
2001cy	  & 52091.44	 & 6.4	& 15.918(0.026)	 & 15.958(0.021)	 & 15.828(0.017)	 & 15.779(0.037) \\ 
2001cy	  & 52092.48	 & 7.5	& 15.988(0.030)	 & 15.809(0.018)	 & 15.817(0.014)	 & 15.738(0.029) \\ 
2001cy	  & 52093.44	 & 8.4	& 15.967(0.042)	 & 15.976(0.036)	 & 15.841(0.044)	 & 15.770(0.034) \\ 
...	  & ...	 & ...	 & ...	 & ...	 & ...	 \\ 
2001do	  & 52136.25	 & 3.1	& 16.127(0.030)	 & 15.924(0.022)	 & 15.719(0.016)	 & 15.618(0.020) \\ 
2001do	  & 52137.27	 & 4.1	& 16.187(0.039)	 & 15.898(0.017)	 & 15.665(0.013)	 & 15.521(0.020) \\ 
2001do	  & 52138.24	 & 5.0	& 16.174(0.031)	 & 15.828(0.017)	 & 15.615(0.016)	 & 15.432(0.016) \\ 
...	  & ...	 & ...	 & ...	 & ...	 & ...	 \\ 
2001fa	  & 52202.35	 & 4.5	& 16.070(0.014)	 & 16.185(0.014)	 & 16.157(0.018)	 & 16.190(0.020) \\ 
2001fa	  & 52203.34	 & 5.4	& 16.012(0.012)	 & 16.122(0.011)	 & 16.070(0.012)	 & 16.039(0.014) \\ 
2001fa	  & 52204.36	 & 6.5	& 16.051(0.013)	 & 16.067(0.010)	 & 15.996(0.016)	 & 15.955(0.011) \\ 
...	  & ...	 & ...	 & ...	 & ...	 & ...	 \\ 
2005dq	  & 53613.31	 & 5.2	& 17.355(0.019)	 & 17.483(0.042)	 & 17.474(0.028)	 & 17.358(0.056) \\ 
2005dq	  & 53614.32	 & 6.2	& 17.358(0.030)	 & 17.400(0.033)	 & 17.399(0.025)	 & 17.355(0.080) \\ 
2005dq	  & 53615.28	 & 7.2	& 17.417(0.023)	 & 17.421(0.039)	 & 17.365(0.022)	 & 17.214(0.059) \\ 
...	  & ...	 & ...	 & ...	 & ...	 & ...	 \\ 
\hline
\end{tabular}\label{t:phot_short}
\tablenotetext{a}{Days since explosion}
\end{table*}

In order to distinguish between SNe~II-L and II-P, we photometrically define an object as a SN~II-L if its $V$-band light curve declines by more than 0.5 mag from peak brightness during the first 50 days after explosion. The division is not very sensitive to this choice (see Section \ref{s:IIPLb}). 

In Figure \ref{f:phot_grid} we show the $BVRI$ light curves of the 11 objects in our sample, marking at the bottom of each plot the epochs at which spectra were taken. Table \ref{t:SNedataP2} includes the redshifts, distances, and explosion dates. Most of the objects are at low redshifts with $z < 0.03$, except for the very luminous SN\,2008es at $z=0.213$ \citep{Miller:2009,Gezari:2009}. The explosion day is set as the midpoint between the first detection and the last non detection. In case a constraining nondetection is not available, as for SN\,2003hk, a rough estimate for the explosion day is obtained by comparing the epoch of peak brightness to that of SNe with similar light curves. Distance measurements are collected from NED\footnote{The NASA/IPAC Extragalactic Database (NED) is operated by the Jet Propulsion Laboratory, California Institute of Technology, under contract with the National Aeronautics and Space Administration (NASA).}, where we use the mean value of distances based on the Tully-Fisher method, Cepheid variable stars, and SNe~Ia.

\begin{table}
	\tiny
	\tabcolsep=0.11cm
\caption{Journal of Spectroscopic Observations}
\begin{tabular}{cccccc}
\hline\hline
SN Name &
MJD &
Age (Days) \tablenotemark{a} &
Instrument &
Range (\AA)&
Exposure (s)\\
\hline
1999co	 & 51368.45	 &36.5	 &Kast	 & 4320-7030 &1200 \\ 
2000dc	 & 51782.25	 &20.4	 &Kast	 & 3290-10450 &900 \\ 
2000dc	 & 51823.18	 &61.4	 &Kast	 & 3300-7790 &1800 \\ 
2000dc	 & 51841.11	 &79.3	 &Kast	 & 3300-7800 &2400 \\ 
2001cy	 & 52106.48	 &21.5	 &Kast	 & 3300-10400 &900 \\ 
2001cy	 & 52117.42	 &32.4	 &Kast	 & 3300-10400 &900 \\ 
2001cy	 & 52144.45	 &59.4	 &Kast	 & 3300-10400 &900 \\ 
2001cy	 & 52164.25	 &79.2	 &Kast	 & 3300-10400 &900 \\ 
2001cy	 & 52172.32	 &87.3	 &Kast	 & 3300-10400 &900 \\ 
2001cy	 & 52207.12	 &122.1	 &Kast	 & 3280-10400 &1500 \\ 
2001do	 & 52144.41	 &11.2	 &Kast	 & 3300-10400 &600 \\ 
2001do	 & 52163.23	 &30.0	 &Kast	 & 3300-10400 &600 \\ 
2001do	 & 52164.19	 &31.0	 &Kast	 & 3300-10400 &600 \\ 
2001do	 & 52172.30	 &39.1	 &Kast	 & 3300-10400 &1000 \\ 
2001fa	 & 52202.25	 &4.3	 &Kast	 & 3300-10400 &900 \\ 
2001fa	 & 52204.00	 &6.1	 &ESI	 & 3930-10198 &300 \\ 
2001fa	 & 52205.00	 &7.1	 &ESI	 & 3930-10199 &200 \\ 
2001fa	 & 52228.39	 &30.5	 &Kast	 & 3280-10400 &600 \\ 
2001fa	 & 52288.00	 &90.1	 &Kast	 & 3268-10570 &2700 \\ 
2003hf	 & 52879.20	 &16.0	 &Kast	 & 3252-10600 &1200 \\ 
2003hf	 & 52885.29	 &22.1	 &Kast	 & 3232-10400 &1500 \\ 
2003hf	 & 52910.00	 &46.8	 &LRIS	 & 3164-9250 &300 \\ 
2003hk	 & 52879.42	 &19.9	 &Kast	 & 3254-10218 &1200 \\ 
2005dq	 & 53615.34	 &7.2	 &Kast	 & 3310-10500 &1800 \\ 
2005dq	 & 53624.38	 &16.3	 &Kast	 & 3350-10400 &1500 \\ 
2005dq	 & 53682.00	 &73.9	 &DEIMOS	 & 4900-10061 &900 \\ 
2007fz	 & 54301.20	 &5.1	 &Kast	 & 3320-10600 &1200 \\ 
2007fz	 & 54319.24	 &23.1	 &Kast	 & 3326-10600 &1500 \\ 
2007fz	 & 54326.00	 &29.9	 &Kast	 & 3420-10130 &1500 \\ 
2007fz	 & 54333.00	 &36.9	 &Kast	 & 3398-10500 &1500 \\ 
2008es	 & 54602.00	 &20.6	 &Kast	 & 3296-10800 &1500 \\ 
2008es	 & 54615.00	 &33.6	 &Kast	 & 3298-10800 &1800 \\ 
2008es	 & 54654.25	 &72.9	 &Kast	 & 3306-10800 &2100 \\ 
2008es	 & 54624.00	 &42.6	 &LRIS	 & 3102-9228 &600 \\ 
2008es	 & 54638.00	 &56.6	 &R.C.S 	 & 3764-9444 &900 \\ 
2008es	 & 54681.00	 &99.6	 &LRIS	 & 3100-9190 &405 \\ 
2008fq	 & 54766.20	 &46.4	 &LRIS	 & 3138-9156 &120 \\ 
2008fq	 & 54739.00	 &19.2	 &Kast	 & 3408-10770 &900 \\ 
2008fq	 & 54746.00	 &26.2	 &Kast	 & 3410-10780 &1200 \\ 
2008fq	 & 54761.00	 &41.2	 &Kast	 & 3428-6404 &2400 \\ 
2008fq	 & 54777.00	 &57.2	 &Kast	 & 3726-10800 &1200 \\ 
2008fq	 & 54790.00	 &70.2	 &Kast	 & 3480-10000 &1500 \\ 
\hline
\end{tabular}
\label{t:spec}
\tablenotetext{a}{Days since explosion}
\end{table}

\section{Photometric Properties}\label{s:phot}

\begin{figure}
\centering
\includegraphics[width=1\columnwidth]{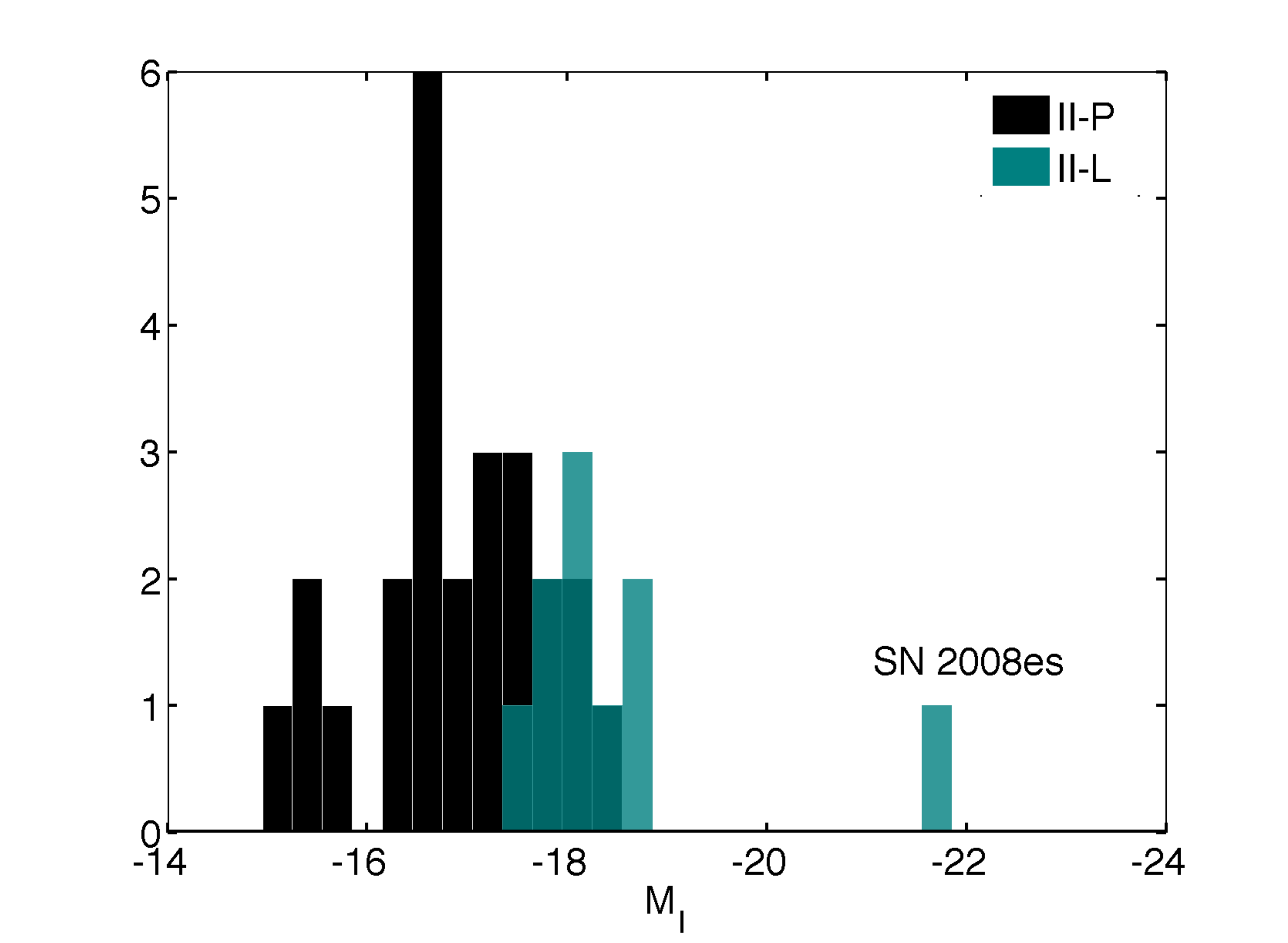}
\caption{$I$-band absolute magnitude light curves. SNe~II-P (black) seem to cover a range of 3 mag (between $-18$ and $-15$ mag) and are $\sim 1.5$ mag dimmer than SNe~II-L (green). }
\label{f:abs_mag}
\end{figure}

Figure \ref{f:abs_mag} presents the luminosity function in the $I$ band, comparing the peak magnitudes of the SNe~II-L to the mean plateau magnitude of SNe~II-P in F14a. Even without the superluminous SN\,2008es, the SNe~II-L are more luminous than SNe~II-P by about 1.5 mag at peak, which has been previously shown in \citet{Anderson:2014} and \citet{Patat:1994}, and more uniform, though the sample is smaller and the subtypes overlap. The data used for this plot are presented in Table \ref{t:data}.

\begin{figure}
\centering
\includegraphics[width=1\columnwidth]{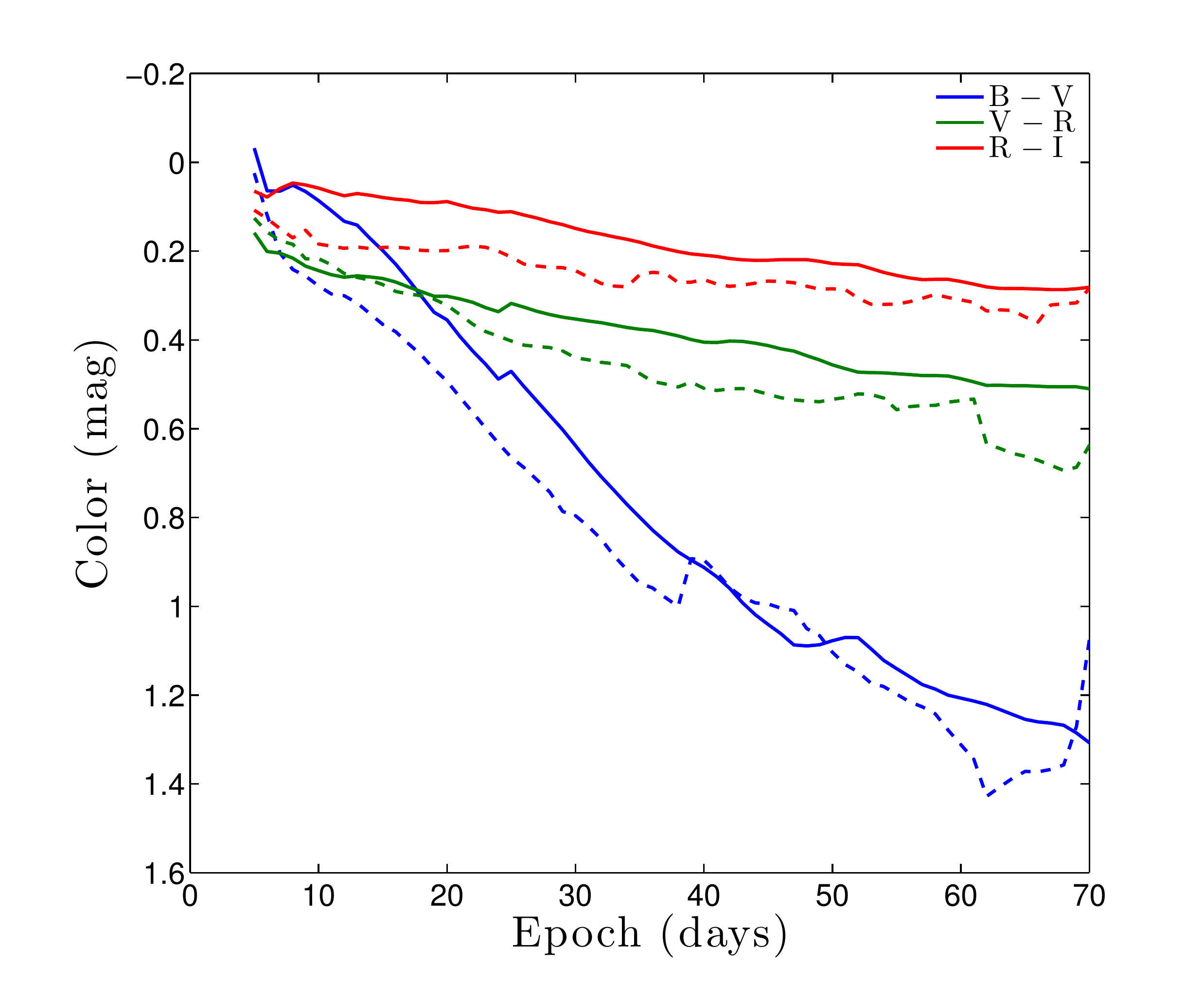}
\caption{Comparison between the mean colour evolution (averaged over different events) of SNe~II-P (solid line) and SNe~II-L (dashed line). The two subtypes evolve similarly in all bands, though SNe~II-L are redder on average than SNe~II-P.}
\label{f:PLcolor_comparison}
\end{figure}

In Figure \ref{f:PLcolor_comparison} we compare the colour evolution of the SNe~II-L to the SNe~II-P from F14a. Aside from a small gap of $\sim0.07$ mag on average (over time), the two subtypes seem to evolve the same way. Since some of the events do not have $R$-band photometry, we only use objects with data in all four bands in order to be consistent when plotting different colours. Therefore, for SNe~II-P we are able to average over a maximum of 11 events per day, and 8 for SNe~II-L. The plot terminates on day 70, since beyond this day the mean SN~II-L colour is determined by only 3~SNe and might be highly affected by individual objects.

The similar evolution of the SN~II-P and II-L colours is not trivial to interpret, since the light curves of the two classes have quite different shapes. The redder colour of SNe~II-L could be intrinsic or caused by a different typical extinction. The latter possibility is consistent with the high mean equivalent width of Na\,I\,D absorption (2.02\,\AA) we find for 5 SNe~II-L, following the methods of F14a. An especially high value of 3.72\,\AA\ is obtained for SN\,2008fq. The mean for the 13 SNe~II-P in F14a, 1.10\,\AA, is significantly lower. If sodium absorption correlates with dust column density \citep{Turatto:2003,Poznanski:2011,Poznanski:2012}, this is an indication that SNe~II-L reside in dustier areas, closer to their birth sites, as one would expect from more massive stars. Alternatively, the Na\,I\,D in SNe~II-L could be partly circumstellar, due to more substantial mass loss.

\begin{table}
	\tiny
	\tabcolsep=0.11cm
\caption{Photometric Properties}
\begin{tabular}{ccccccc}
\hline\hline
SN Name &
M$_{\rm I,max}$ &
Decline ($I$; mag/50\,d) &
Decline ($V$; mag/50\,d) \\
\hline
1999bg	 &-16.61(0.40)	&0.11(0.02)	&0.33(0.01)	\\ 
1999co	 &-18.24(0.22)	&0.38(0.09)	&0.77(0.05)	\\ 
1999d	 &-   	&-   	&-   	\\ 
1999em	 &-16.75(0.36)	&-0.13(0.01)	&0.19(0.00)	\\ 
1999gi	 &-15.61(0.31)	&-0.02(0.01)	&0.36(0.02)	\\ 
2000bs	 &-17.83(0.22)	&-0.02(0.04)	&0.30(0.02)	\\ 
2000dc	 &-18.02(0.31)	&0.62(0.04)	&0.94(0.13)	\\ 
2000dj	 &-17.26(0.22)	&0.06(0.03)	&0.38(0.03)	\\ 
2001bq	 &-17.35(0.22)	&-0.08(0.02)	&0.24(0.02)	\\ 
2001cm	 &-17.25(0.37)	&-0.07(0.02)	&0.37(0.03)	\\ 
2001cy	 &-18.20(0.22)	&0.35(0.01)	&0.82(0.01)	\\ 
2001do	 &-17.95(0.22)	&0.72(0.02)	&1.25(0.02)	\\ 
2001fa	 &-18.75(0.01)	&1.41(0.03)	&1.99(0.03)	\\ 
2001hg	 &-16.81 0.29 	&-   	&-   	\\ 
2001x	 &-16.64(0.28)	&-0.04(0.01)	&0.21(0.01)	\\ 
2002an	 &-17.83(0.05)	&-	&-	\\ 
2002bx	 &-18.42 0.42 	&-   	&-   	\\ 
2002ca	 &-16.27(0.22)	&0.05(0.03)	&0.39(0.03)	\\ 
2002gd	 &-16.39(0.21)	&0.11(0.02)	&0.25(0.05)	\\ 
2002hh	 &-15.32(0.19)	&0.01(0.05)	&0.31(0.06)	\\ 
2003gd	 &-16.51 0.23 	&-   	&-   	\\ 
2003hf	 &-18.58(0.22)	&1.24(0.06)	&2.00(0.04)	\\ 
2003hk	 &-18.41(0.06)	&0.79(0.02)	&1.23(0.02)	\\ 
2003hl	 &-17.17(0.30)	&0.03(0.01)	&0.37(0.01)	\\ 
2003iq	 &-17.57(0.30)	&0.04(0.01)	&0.37(0.01)	\\ 
2003z	 &-15.20(0.20)	&-0.26(0.03)	&0.19(0.02)	\\ 
2004du	 &-18.07(0.33)	&0.09(0.02)	&0.48(0.02)	\\ 
2004et	 &-17.53(0.19)	&-0.05(0.00)	&0.36(0.00)	\\ 
2005ay	 &-16.52(0.40)	&-0.14(0.01)	&0.19(0.01)	\\ 
2005cs	 &-15.52(0.07)	&- &0.26(0.01)	\\ 
2005dq	 &-17.72(0.22)	&0.85(0.05)	&1.32(0.05)	\\ 
2007fz	 &-17.50(0.22)	&1.40(0.05)	&2.55(0.08)	\\ 
2008es	 &-21.75(0.22)	&1.06(0.17)	&1.44(0.11)	\\ 
2008fq	 &-18.04(0.36)	&0.81(0.73)	&2.34(4.80)	\\ 
\hline
\label{t:data}
\end{tabular}
\end{table}

\begin{figure}
\centering
\includegraphics[width=1\columnwidth]{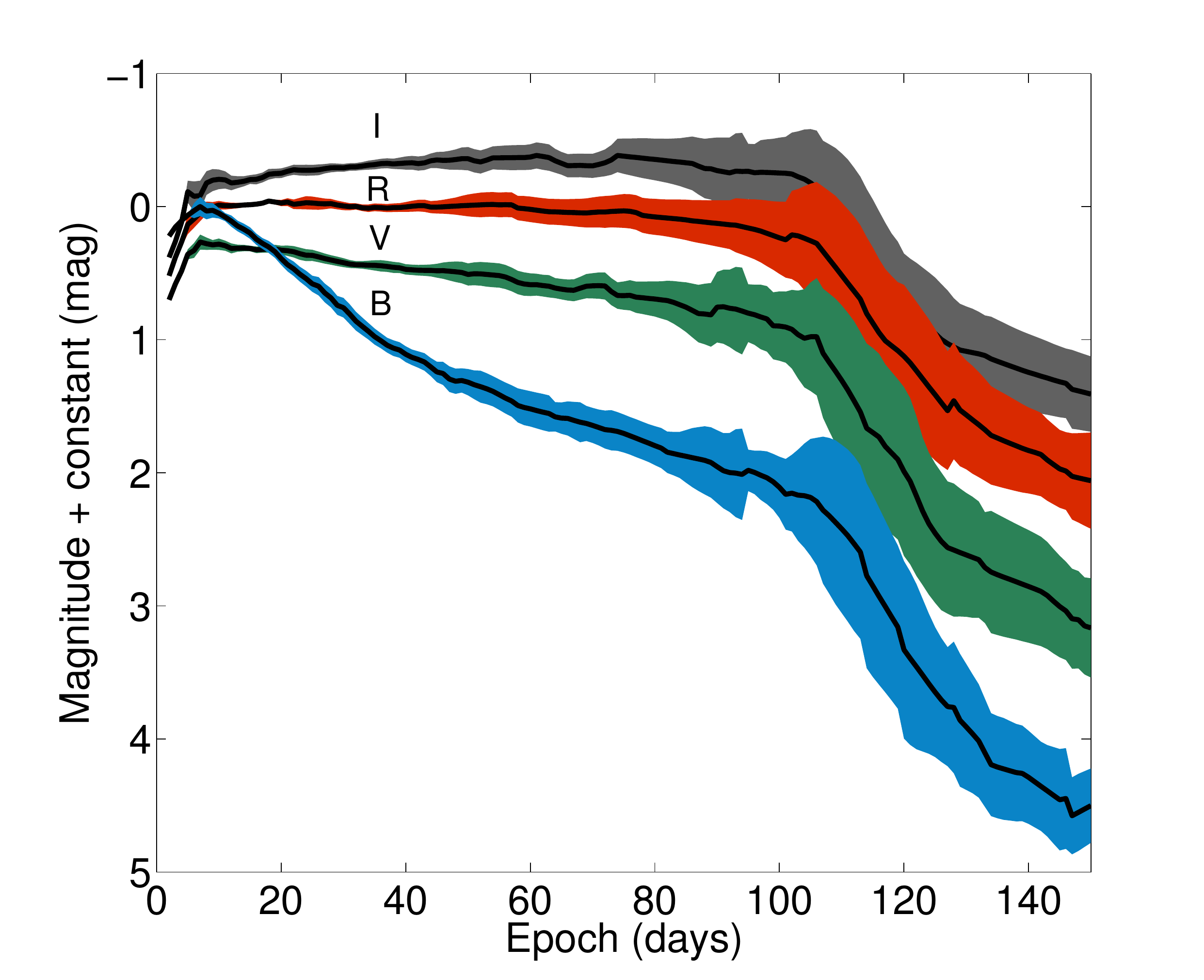}
\includegraphics[width=1\columnwidth]{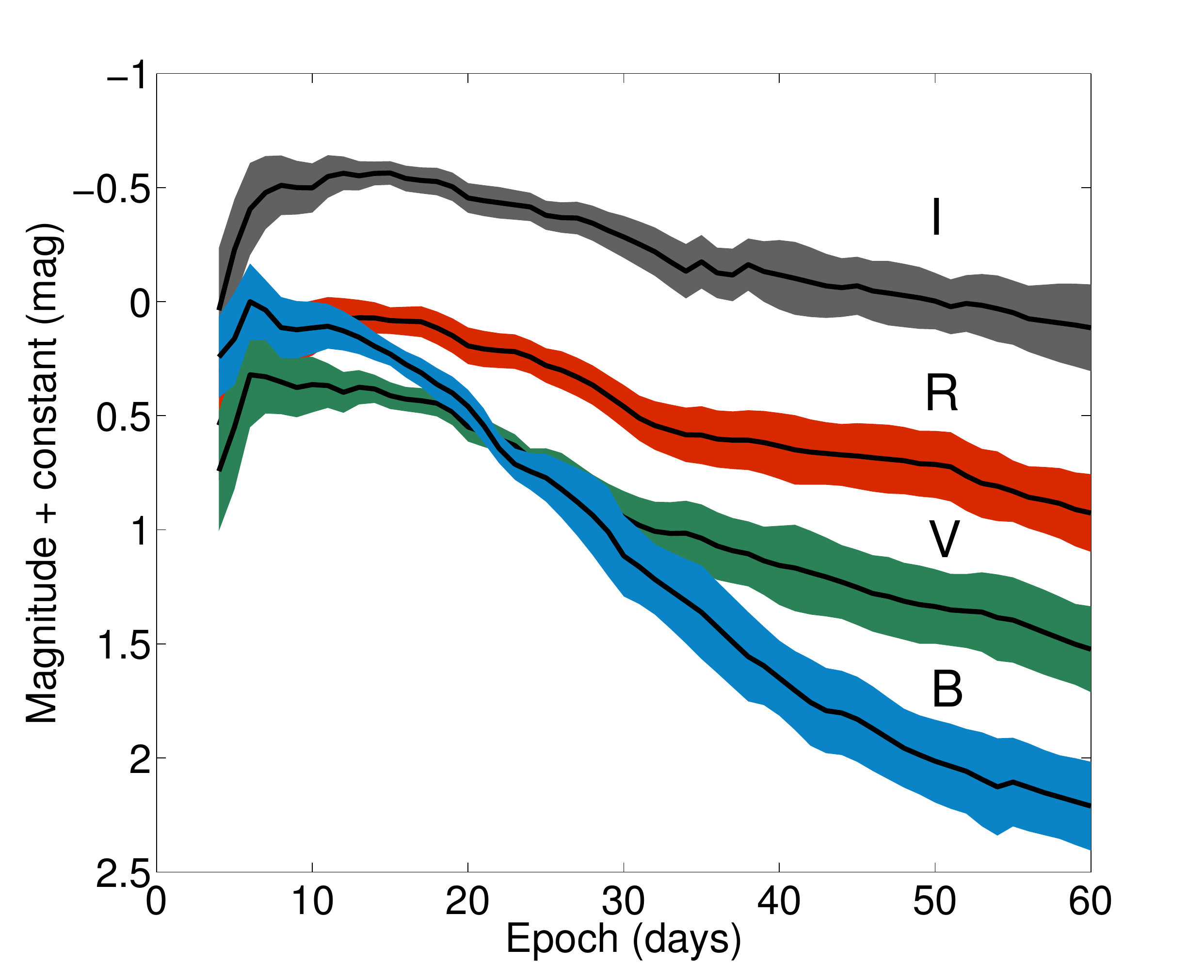}
\caption{Template SN~II-P (top) and SN~II-L (bottom) light curves produced using our sample. The curves are offset to have zero $B$-band magnitude at peak and the colours found in Section \ref{s:phot}.}
\label{f:tmpIIP}

\end{figure}

We produce median light curves (in 4 bands, as opposed to the R-band only templates in \citealt{Li:2011}) for the two SN subtypes as follows. Every SN, in every band, was offset in magnitude, so as to minimise scatter before day 50 for a given SN subtype. We then interpolate the light curves linearly in time and calculate the median and median absolute deviation using all available objects. This gives reasonably tight results before day 60 for SNe~II-L and day 150 for SNe~II-P. The templates are then offset to peak at magnitude zero in $B$ and have at peak the colours we find above. The result can be seen in Figure \ref{f:tmpIIP} and is downloadable from \url{http://www.astro.tau.ac.il/~dovip/IIPIIL_templates.txt}.

\subsection{SN~II-P -- II-L -- IIb -- A Continuum?}
\label{s:IIPLb}

SNe~II-P go through a prolonged phase of hydrogen recombination in the envelope, forming a plateau-shaped light curve followed by a sharp decline to the nebular phase. SNe of Type II-L, on the other hand, decline linearly (in magnitudes) after reaching their luminosity peak, $\sim 10$ days after the explosion. The lack of a plateau could suggest that the progenitor has shed a large fraction of its mass before the explosion, so that at the end of its life it has a low-mass hydrogen envelope which cannot sustain a long recombination phase. One could therefore expect a continuum of plateau durations and decline rates, corresponding to different envelope masses. 

Based on a sample of 15 SNe~II, \citet{Arcavi:2012} suggested that there is a clear photometric division of the various SNe~II into 3 subclasses rather than a continuum of light-curve shapes. This may indicate that Type II-P, II-L, and IIb events result from distinct progenitor systems (instead of, for example, a one-dimensional mass sequence) such that some thresholds (in binarity or metallicity) guide the stellar evolution into different regimes. More recently, \citet{Anderson:2014} and \citet{Sanders:2014} published two large samples of SNe~II and find that there is no clear gap in the distributions,  that there is no obvious photometric threshold that separates the SNe~II-P from the SNe~II-L.

\begin{figure}
\centering
\includegraphics[width=1\columnwidth]{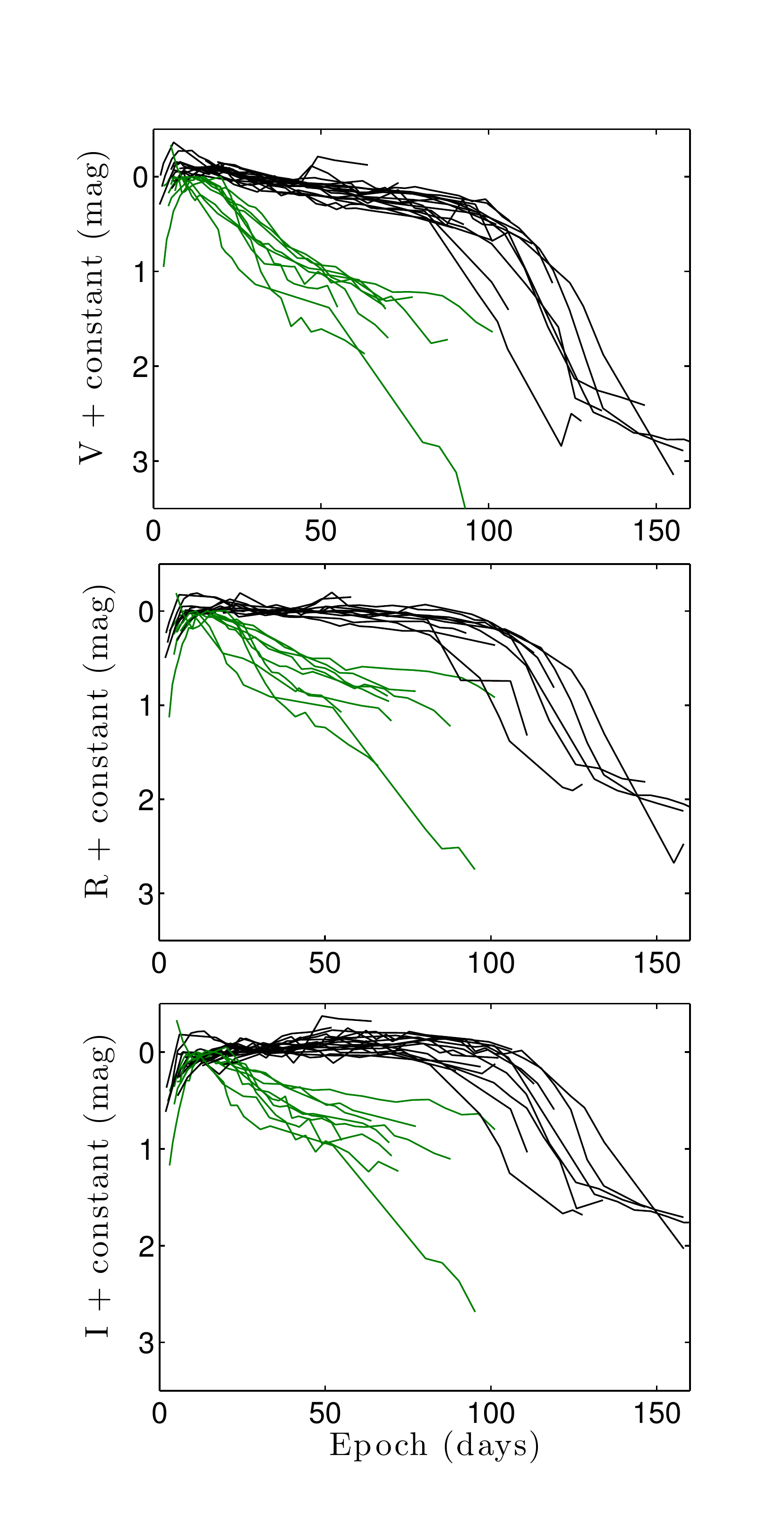}
\caption{Shifted light curves of all events in the $BVRI$ bands. The curves are visually classified into two subclasses according to the shape of their light curves: SNe~II-P (black curves) and linearly declining SNe~II-L (green curves). The subclasses seem to be clearly separated by a gap in the $V$, $R$, and $I$ bands.}
\label{f:compare_light_curves}
\end{figure}

In order to test that picture, we repeat the process done by \citet{Arcavi:2012}. Our SN~II-L and SN~II-P light curves in $V$, $R$, and $I$ are plotted in Figure \ref{f:compare_light_curves}. SNe~II-L are shifted to have the same peak magnitude, while SNe~II-P are shifted by the average plateau magnitude in the $I$ band during the first 50 days after explosion. This method introduces a bias, since in effect we are assigning the classes before we plot the light curves. However, this would only influence the appearance of the division, as we measure decline self consistently for all SNe. Since it is hard to define a plateau in the $B$ band for SNe~II-P, we exclude this band from the analysis. 

Our results partly reproduce the findings of \citet{Arcavi:2012}, who did the same analysis with the $R$-band photometry of 15 SNe~II. In all bands the plateau and declining curves do not form a continuum but cluster into two clearly separated subclasses. As discussed by F14a, the SNe~II-P show a uniformity of plateau durations of $\sim 100$ days, while SNe~II-L span a range of decline rates.

\begin{figure}
\centering
\includegraphics[width=1\columnwidth]{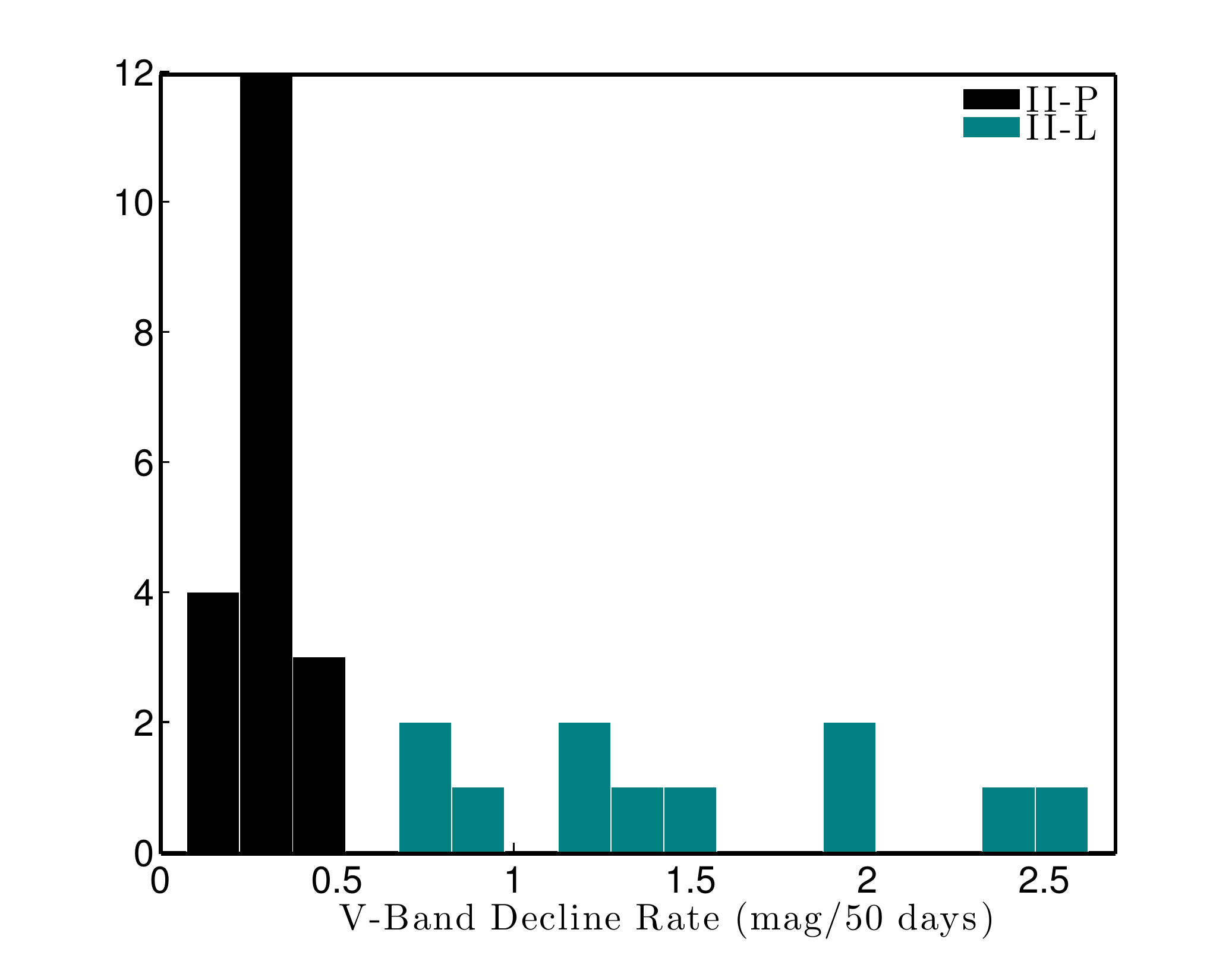}
\caption{$V$-band decline distribution. We define as SNe~II-P (in black) those that decline by less than 0.5~mag from peak brightness in the first 50 days after explosion. Faster decliners are the SNe~II-L (shown in green). }
\label{f:decline_hist}
\end{figure}

\begin{figure}
\centering
\includegraphics[width=1\columnwidth]{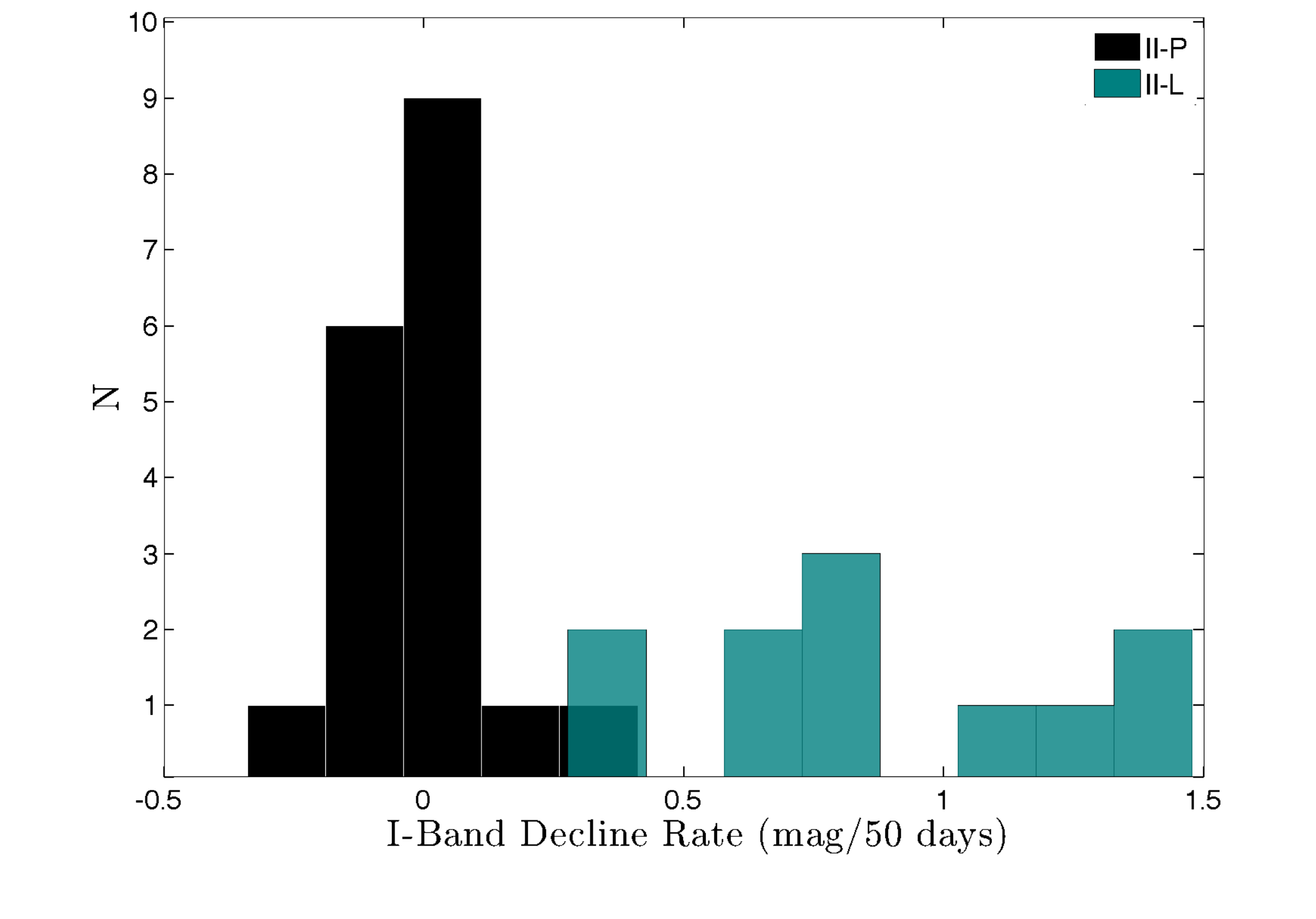}
\caption{$I$-band decline distribution. Classification is based on the decline in V-band (see figure \ref{f:decline_hist}). }
\label{f:decline_histI}
\end{figure}

However, in figures \ref{f:decline_hist} and \ref{f:decline_histI} we show the distribution in decline rates in bands $V$ and $I$ respectively, of our SN~II-P and SN~II-L samples. Clearly, now the classes are not trivial to separate, showing that the methodology of \citet{Arcavi:2012} was indeed introducing a bias. Nevertheless, while our small sample is unimodal, it has a long tail which is no very characteristic of a single population (the decline rate values are given in Table \ref{t:data}). We add to this plot the decline rate values calculated by \citet{Anderson:2014} for a large sample of SNe. \citet{Anderson:2014} defined the parameter `$s2$' to account for the decline rate of the light curve during the recombination period, following the early bump. This definition is similar, though not identical, to the decline rate values of the plateau presented in this plot. The maximum of our distribution falls between the two peaks of \citet{Anderson:2014}'s sample. However, The decline rate distribution of the SNe from \citet{Anderson:2014} peaks around 0.55\,mag$/50$\,d, a value we would consider rather II-L-like. Their distribution actually seems bimodal, with a large fraction of SNe with no significant decline, and another with a broad range of decline rates, though this may be sensitive to binning.

\begin{figure}
\centering
\includegraphics[width=1\columnwidth]{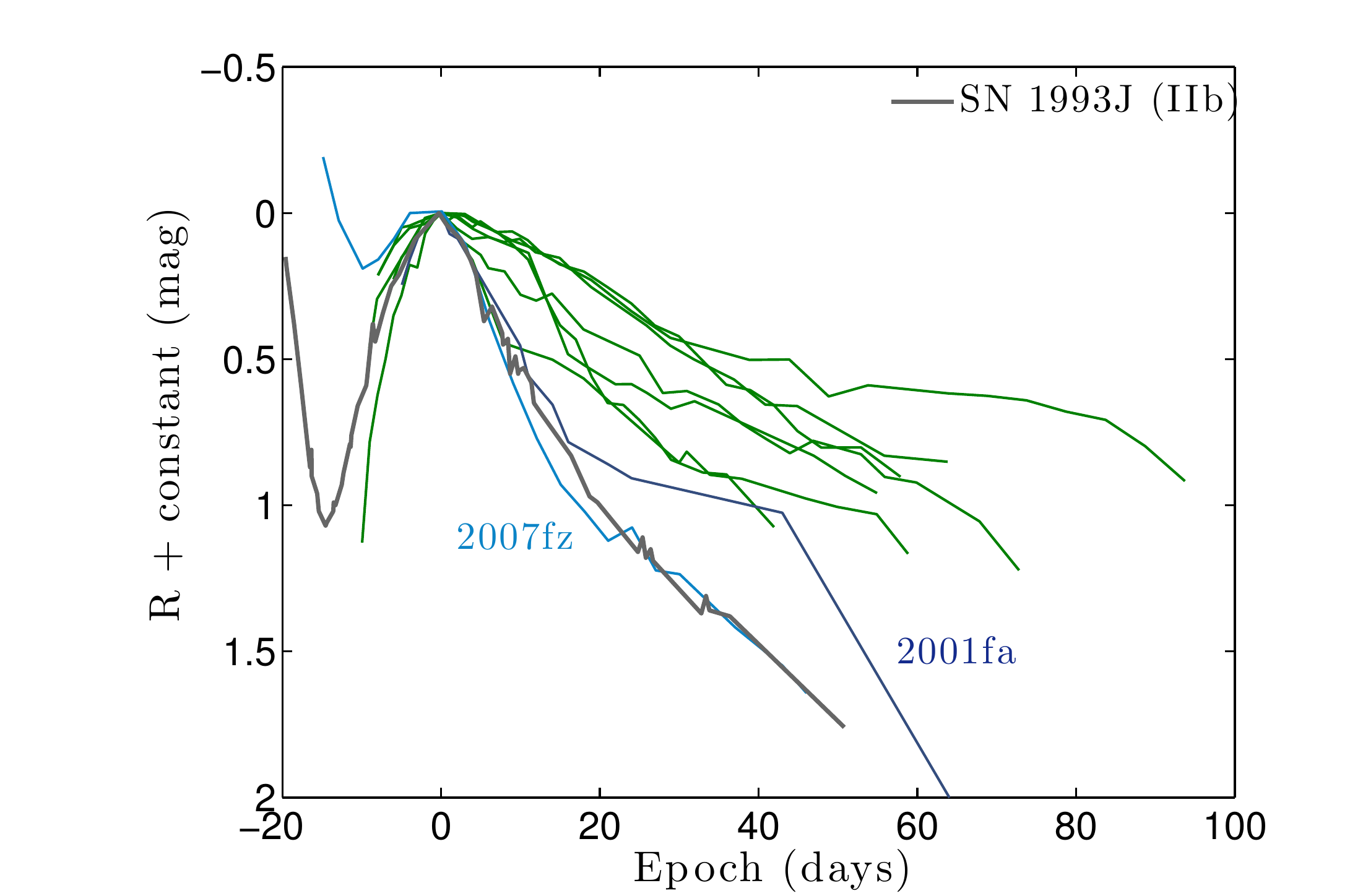}
\caption{$R$-band photometry of SNe~II-L, with the Type IIb SN\,1993J for comparison. The light curves are aligned in magnitude and in time by their rise time to peak luminosity. Photometrically, the SN~IIb and SN~II-L subclasses seem to form a continuum.}
\label{f:IIL_lc_continuum}
\end{figure}

To compare the SN~II-L subclass to SNe~IIb, we compare our light curves with those of the classic Type IIb SN\,1993J (data taken from \citealt{Richmond:1994}). Aside from its late rise to peak brightness, this SN~IIb seems to resemble some of the SNe~II-L (e.g., SNe\,2007fz and 2001fa) in its general shape and decline rate. This similarity is apparent in Figure \ref{f:IIL_lc_continuum}, where we set the $R$-band maximum of the SNe~II-L to day zero (all events except SN\,2001cy and SN\,2008es have data at peak brightness). We choose this reference day owing to the variations in rise time. Relative to this reference point, it is clear that SN\,1993J is merely a continuation of the SN~II-L subclass, for which it serves as a lower bound with the highest decline rate. This behaviour is consistent in the $V$ and $I$ bands as well. This is contrary to the indications found by \citet{Arcavi:2012} that the SN~IIb light curves are well separated from those of SNe~II-L. 

One should wonder whether our sample is indeed clearly free of SNe~IIb. The light curves of SN\,2007fz and SN\,2001fa  bear a striking resemblance to that of SN\,1993J; indeed, SN\,2007fz even shows the initial decline and rise to peak magnitude. We examine and mostly reject this hypothesis in Section \ref{s:spec}, where we show that all our SNe seem to be more spectroscopically consistent with SNe~II-L rather than with SNe~IIb.

SNe~II-L seem to form a photometric class that might be distinct from SNe~II-P, but similar to SNe~IIb during their decline. Since the rise time is in effect a tracer of the radius of the star when exploding, while the later parts of the light curve probe the physics of the expanded ejecta, this may indicate that SNe~IIb and SNe~II-L come from different progenitor channels, but both have been stripped of a significant fraction of their hydrogen via some (perhaps different) mechanism. As we show below, spectroscopy further supports this scenario.

\section{Spectroscopic Properties}\label{s:spec}

\begin{figure*}
\centering
\includegraphics[width=0.85\textwidth]{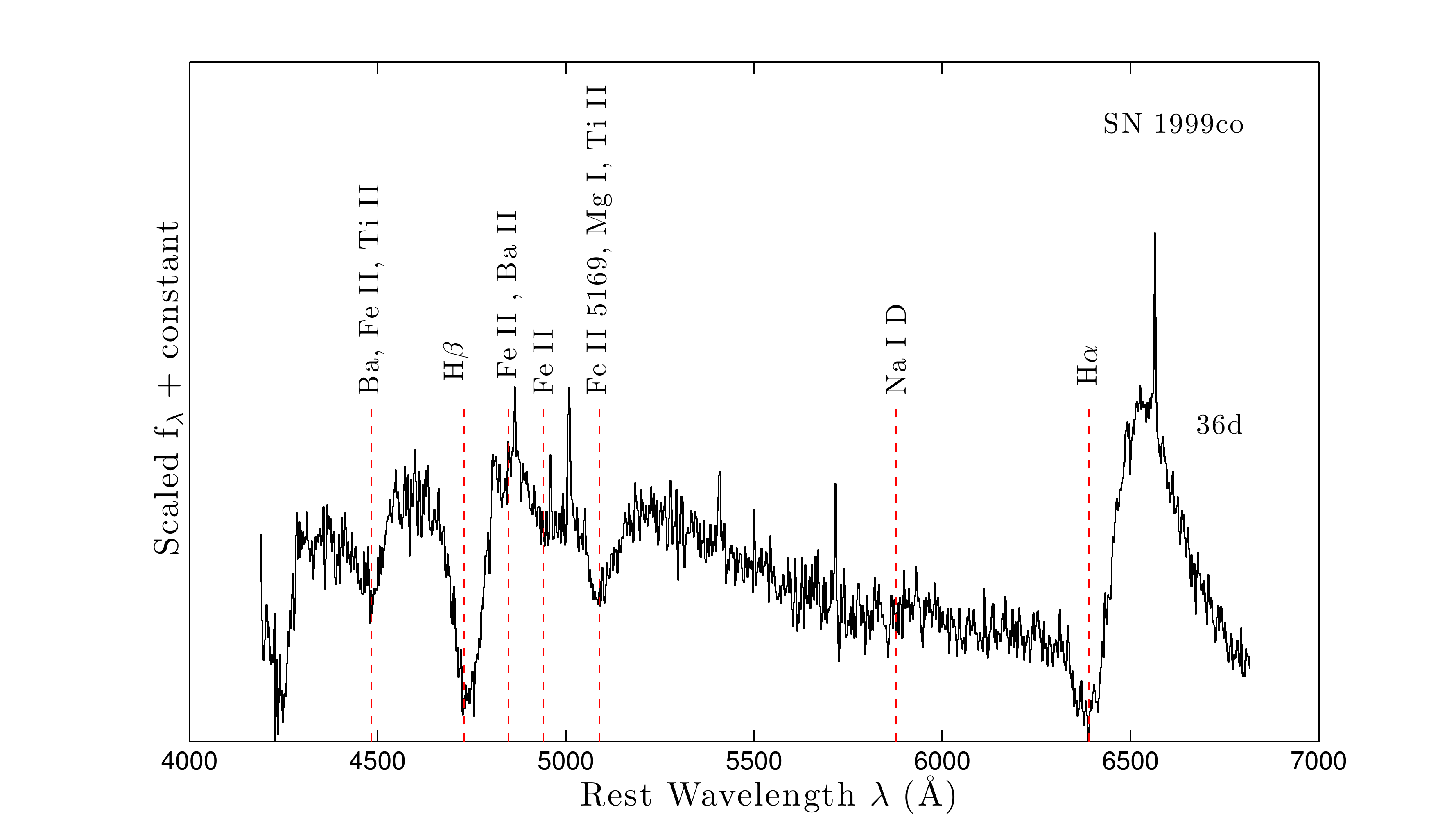}
\includegraphics[width=0.85\textwidth]{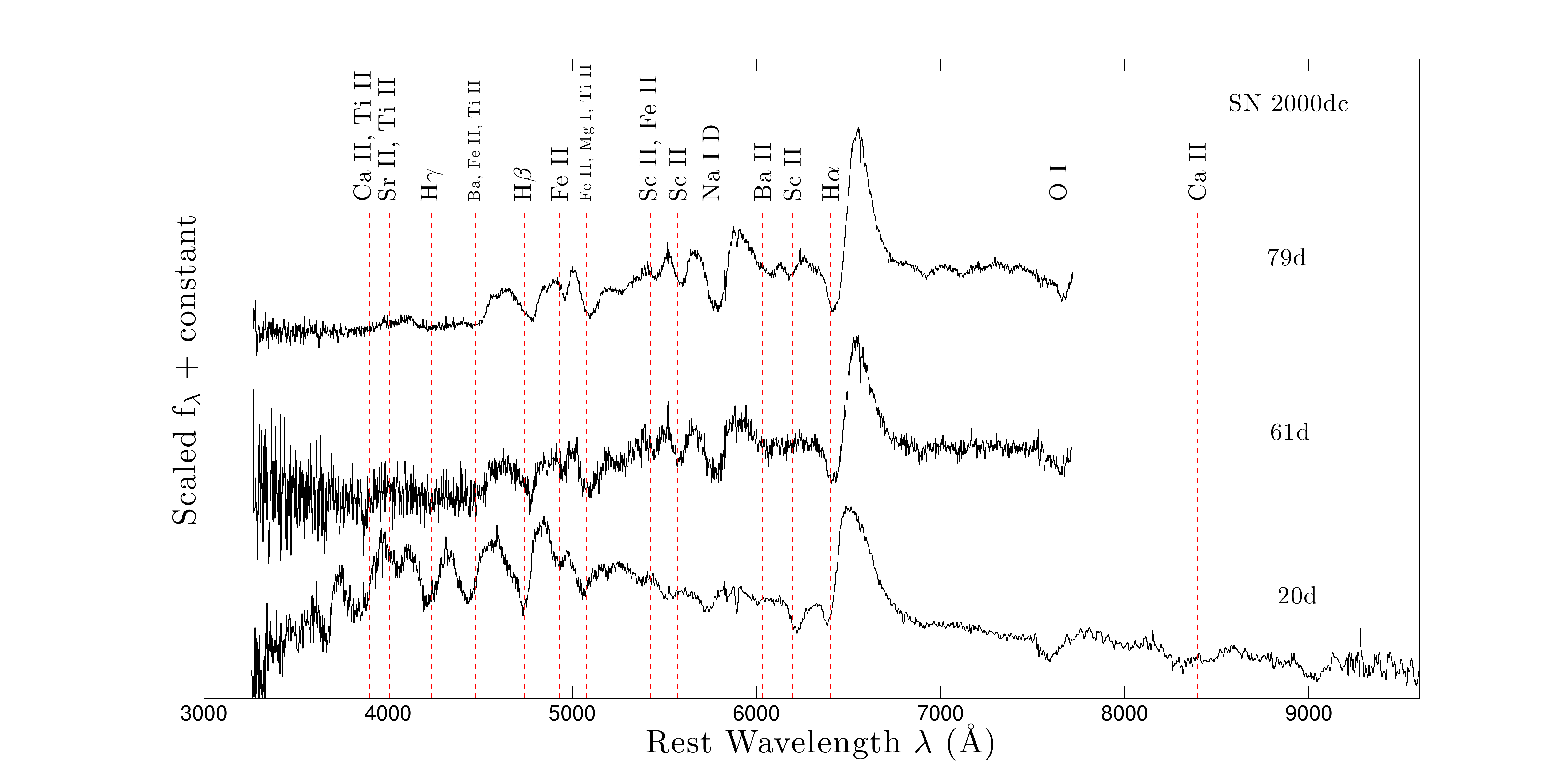}
\includegraphics[width=0.85\textwidth]{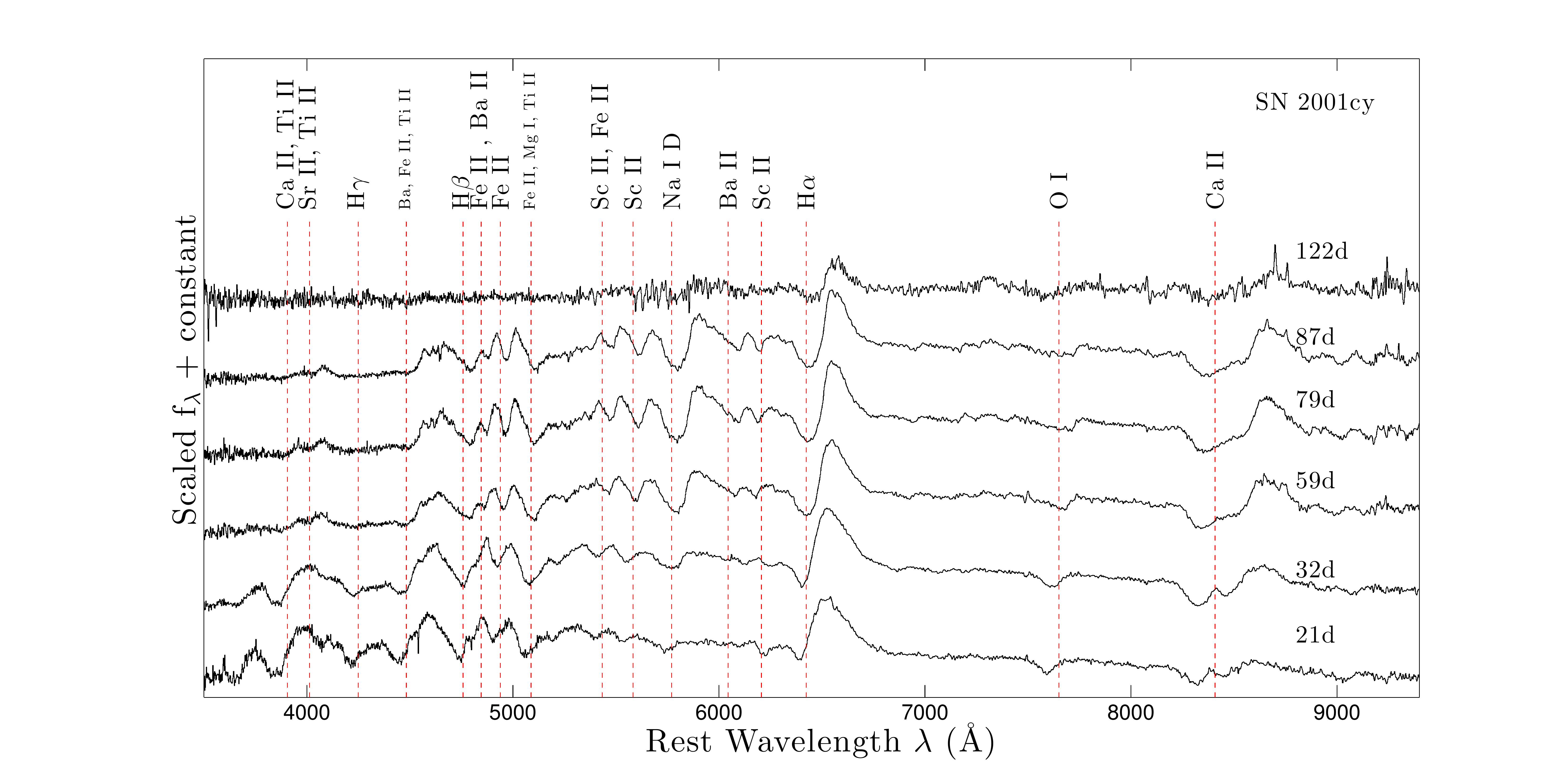}
\caption{Spectra of SNe\,1999co, 2000dc, and 2001cy. Spectral lines appear underdeveloped during early epochs, compared to typical SNe\,II-P, and show distinct features only starting 20--30 days after explosion. The spectra also seem to lack strong absorption of hydrogen, particularly H$\alpha$. Metal lines are stronger compared to H$\alpha$, evident in the absorption profiles of O\,I $\lambda$7774 and Na\,I\,D/He\,I $\lambda$5876.}\label{f:s2L1}
\end{figure*}

\begin{figure*}
\centering
\includegraphics[width=0.9\textwidth]{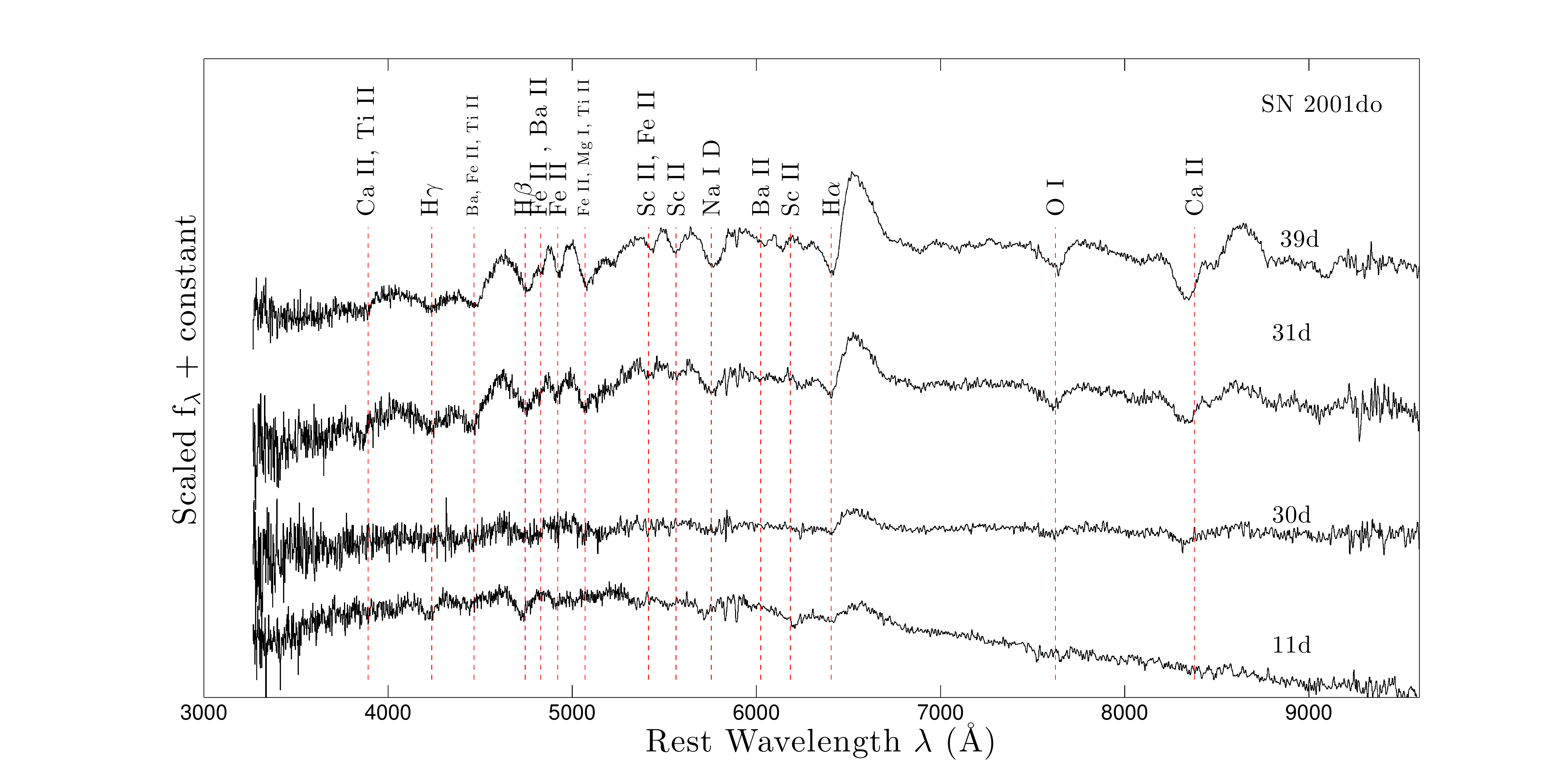}
\includegraphics[width=0.9\textwidth]{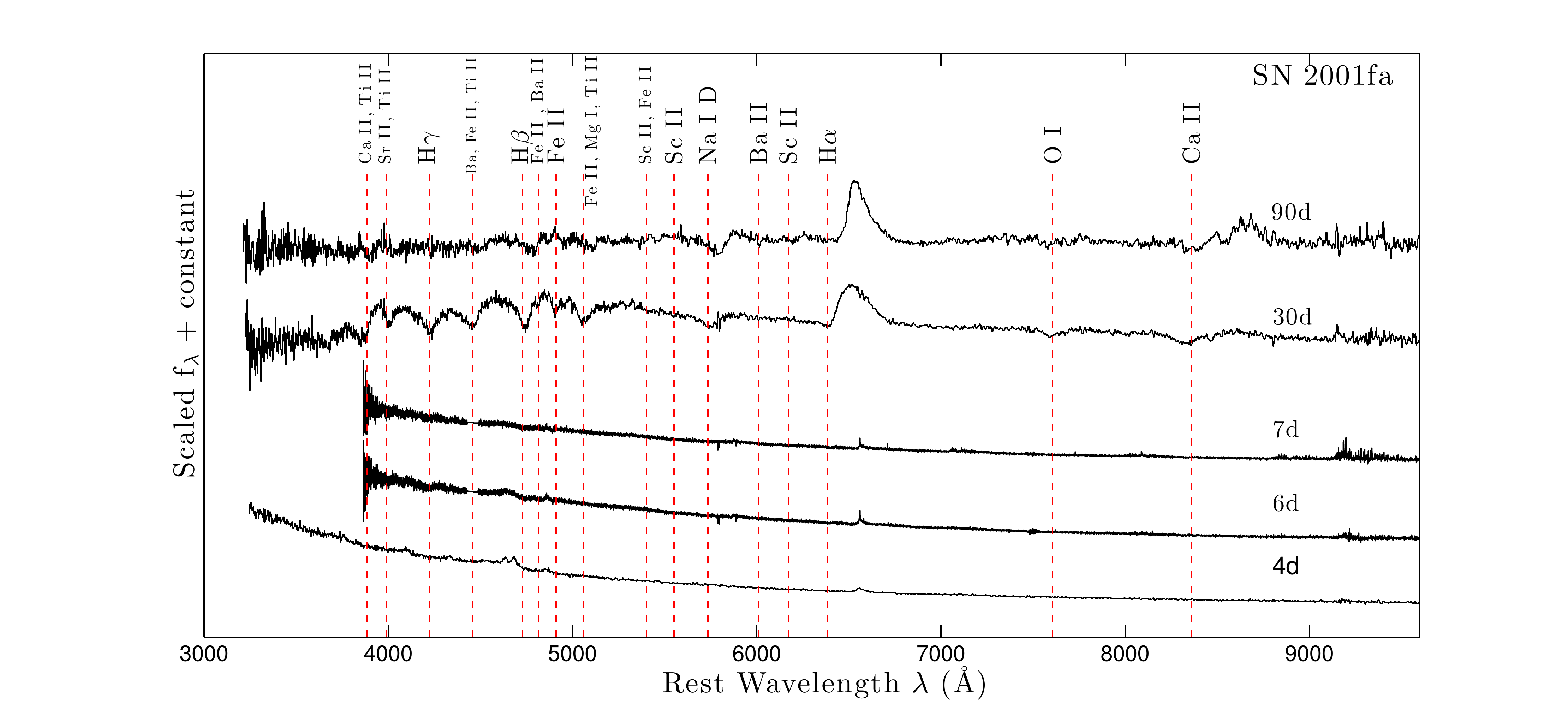}
\includegraphics[width=0.9\textwidth]{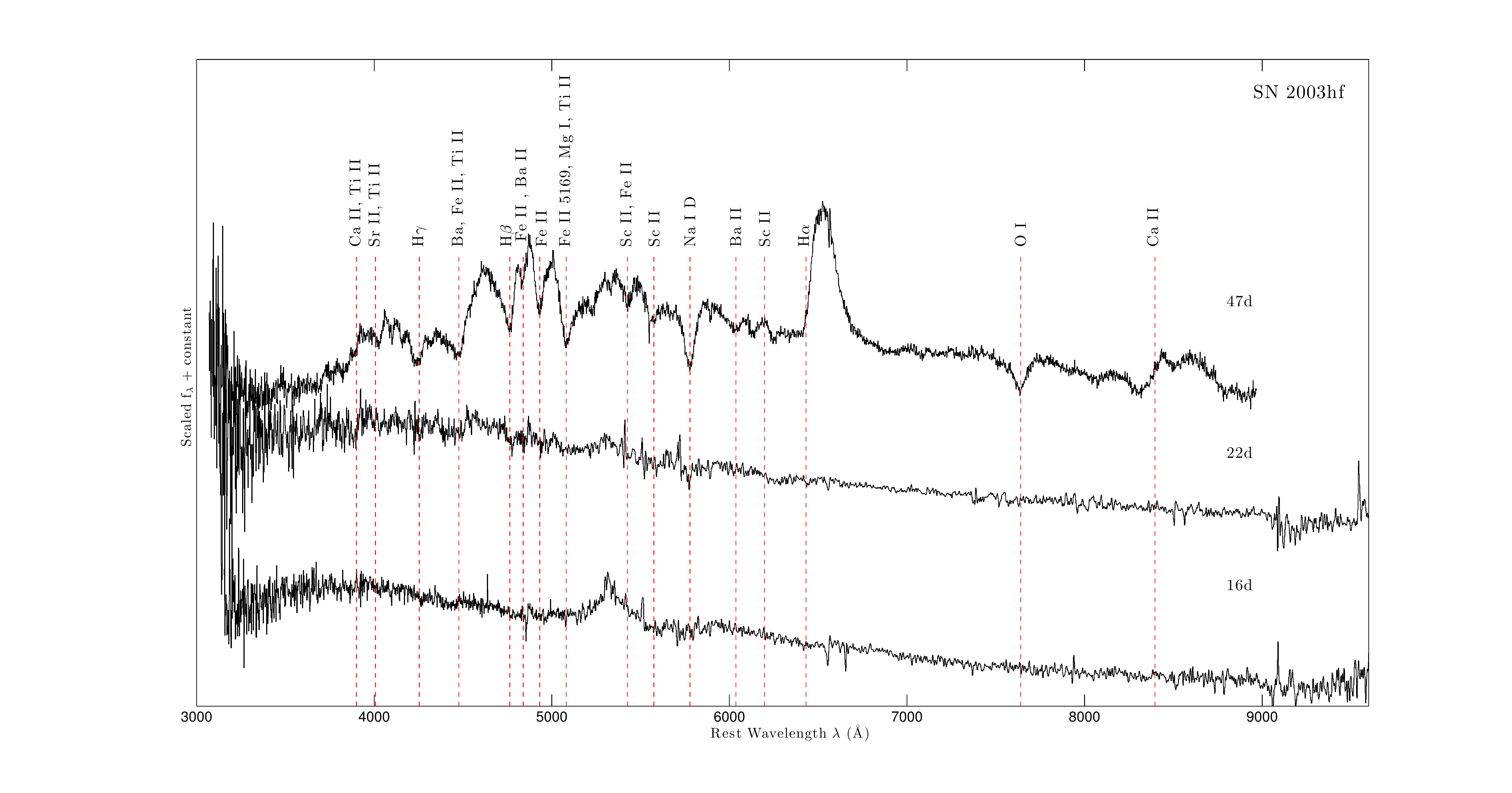}
\caption{Same as Figure \ref{f:s2L1} for SNe\,2001do, 2001fa, and 2003hf.}\label{f:s2L2}
\end{figure*}

\begin{figure*}
\centering
\includegraphics[width=0.9\textwidth]{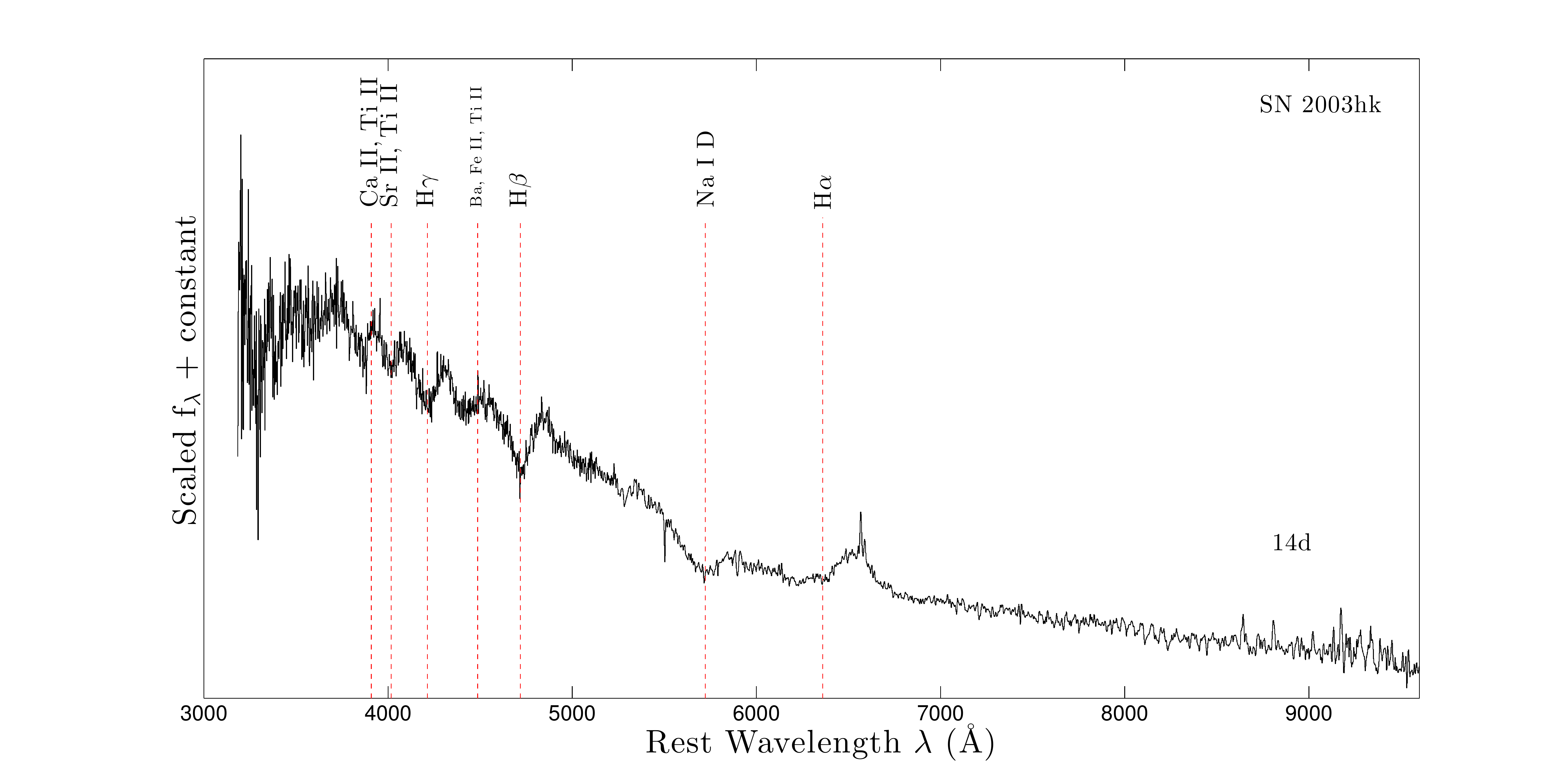}
\includegraphics[width=0.9\textwidth]{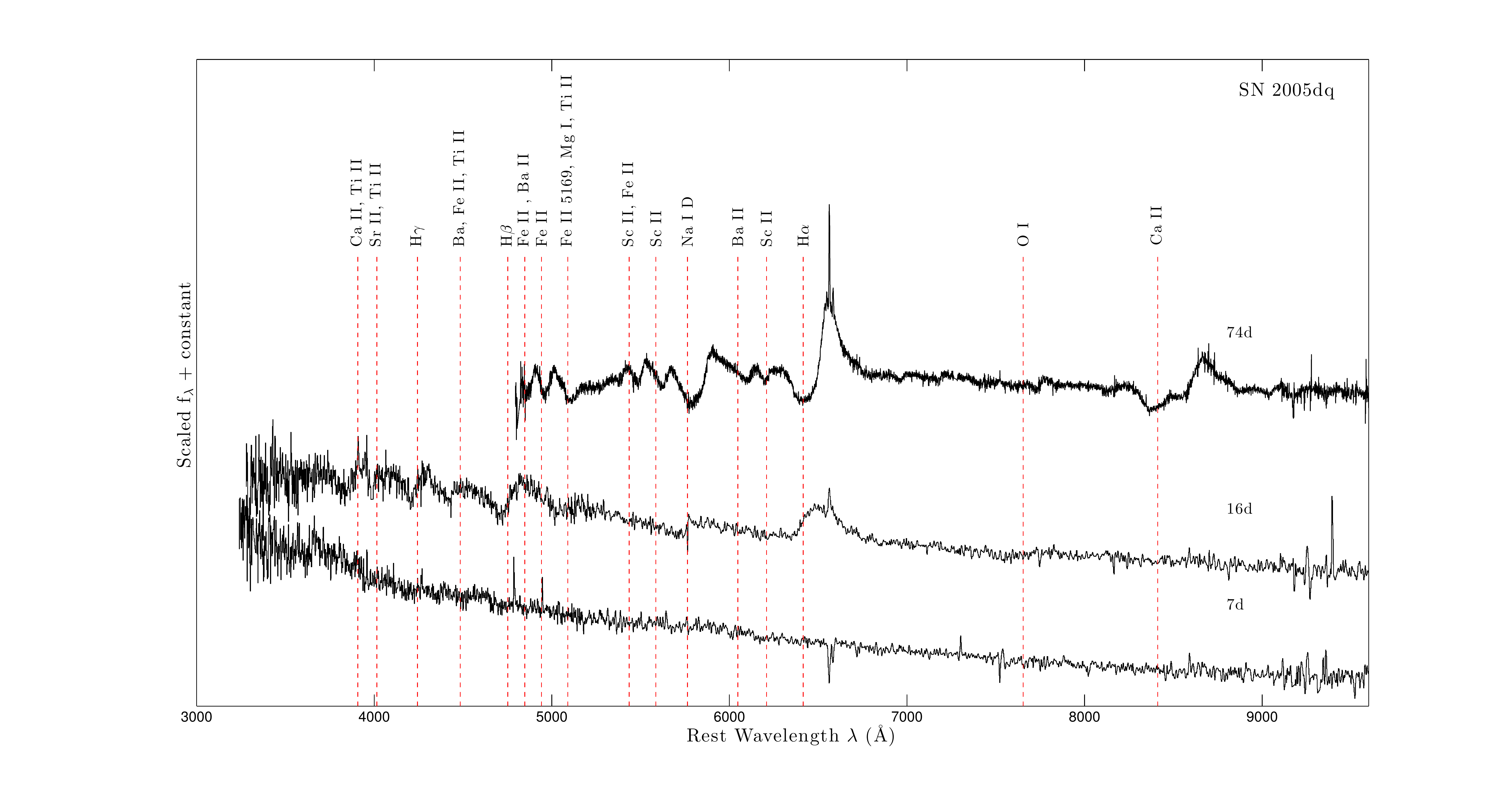}
\includegraphics[width=0.9\textwidth]{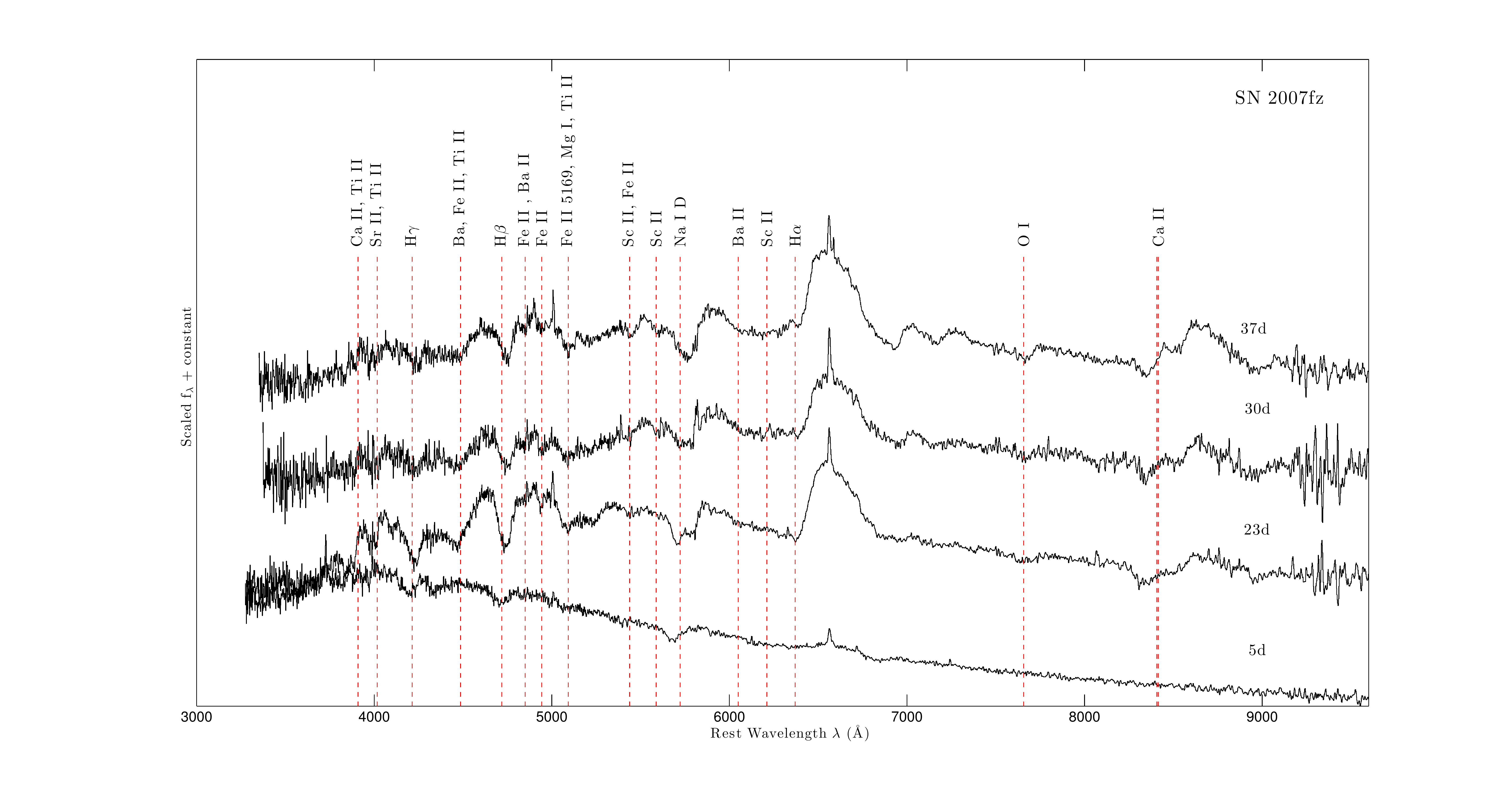}
\caption{Same as Figure \ref{f:s2L1} for SNe\,2003hk, 2005dq, and 2007fz.}\label{f:s2L3}
\end{figure*}

\begin{figure*}
\centering
\includegraphics[width=0.9\textwidth]{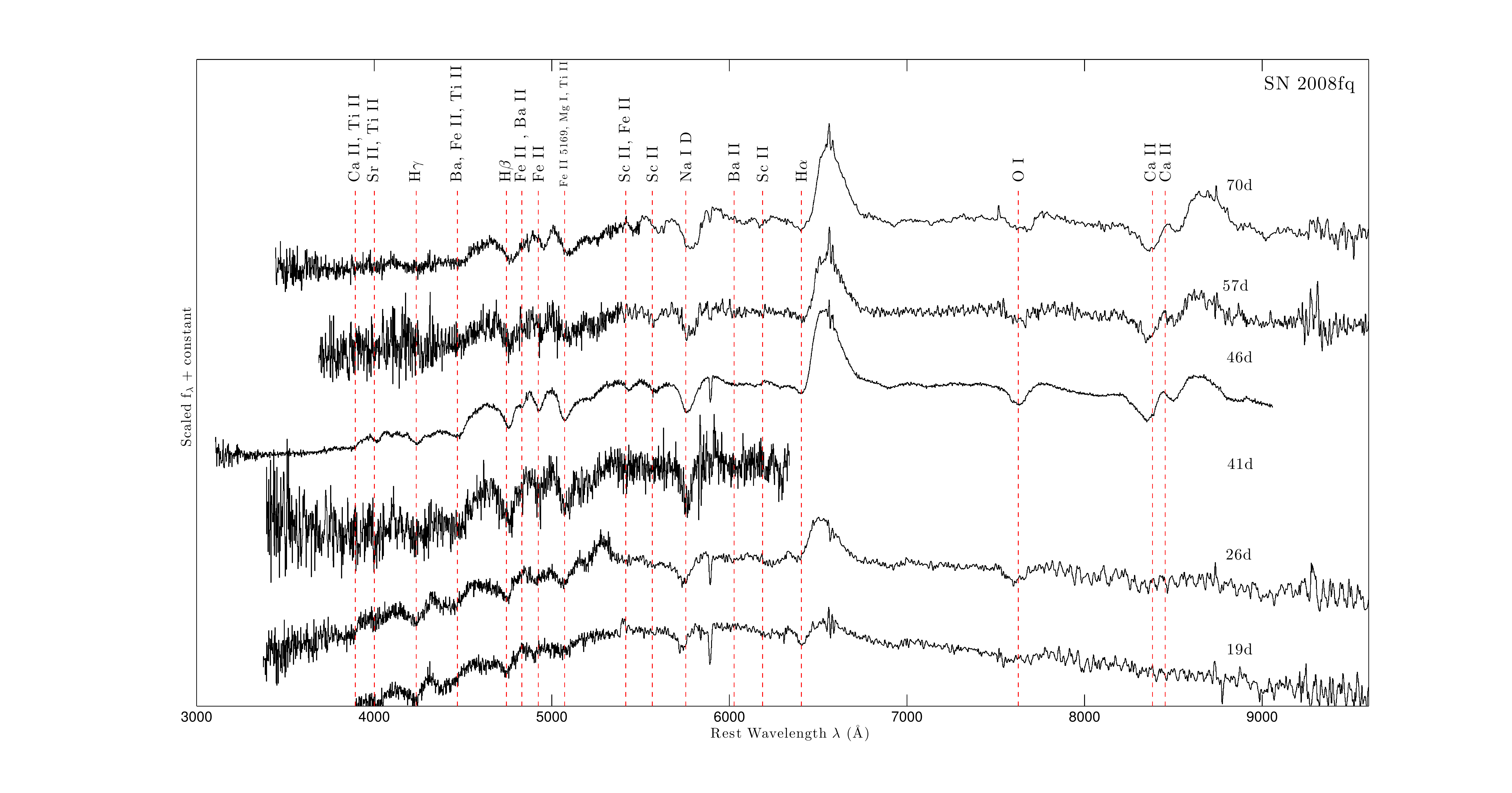}
\includegraphics[width=0.9\textwidth]{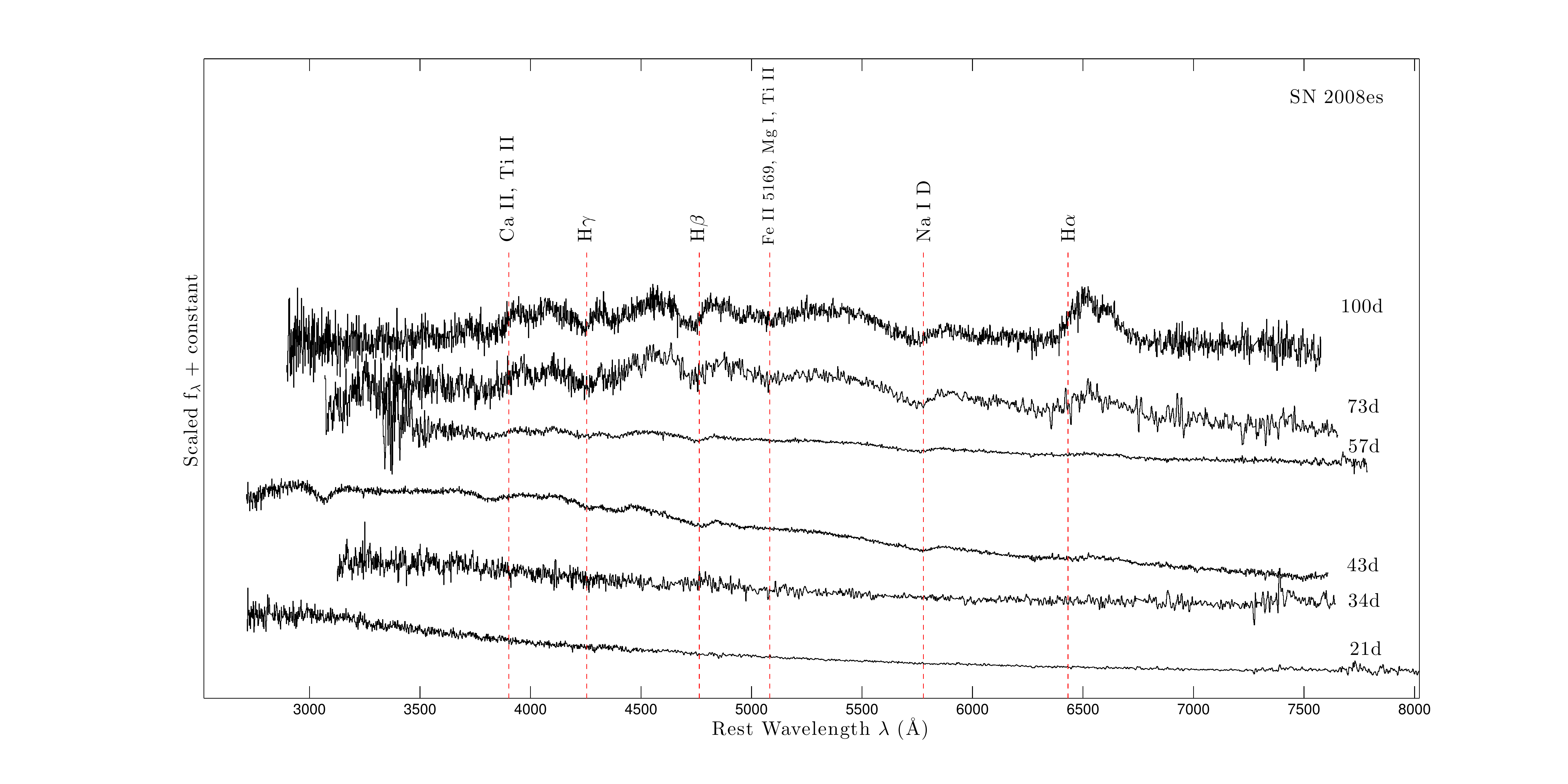}
\caption{Same as Figure \ref{f:s2L1} for SNe\,2008fq and 2008es.}\label{f:s2L4}
\end{figure*}

Figures \ref{f:s2L1}--\ref{f:s2L4} show optical spectra of all SNe~II-L in our sample, with spectral lines identified by using the synthetic spectra from \citet{Hatano:1999}, as described in detail by F14a.

In contrast with standard SN~II-P spectra, in SN~II-L spectra the hydrogen lines only typically begin to appear at later epochs and are usually weaker than in SNe II-P, as would be expected for hydrogen-poor envelopes. 

SN\,2001fa, for example, has very few features during its first 9 days. The spectrum on day 30 contains H$\alpha$ in emission, yet with very little absorption, and a much stronger H$\beta$ line. SN\,2003hf displays no features at all at day 16, except for a strange emission feature (possibly an artifact?) near 5350\,\AA, which then almost disappears in the subsequent spectrum on day 22 (see more below). There is no sign whatsoever of hydrogen in the spectra in the first three weeks, and on day 47 we do see strong H$\alpha$ emission but no absorption. Most other objects seem to behave in a similar fashion. 

The early-time spectra of SNe\,2001fa and 2007fz have narrow H$\alpha$ emission lines, perhaps from circumstellar interaction, but SN\,2001fa also shows the C\,III and N\,III lines recently identified by \citet{gal-yam14} as being emitted from the outer envelope of the dying star, before it is destroyed by the ejecta. These signs disappear very early in the evolution of the SN, making it hard to determine whether they are present only in the spectra of those two rapidly declining objects or in the rest of the Type II-L subsample as well, for which we do not have such early-time spectra. The emission feature in the first spectrum of SN\,2003hf is probably too blue to be He\,II 5411\,\AA\ as in SN\,2013cu \citep{gal-yam14}. 

\begin{figure*}
\centering
\includegraphics[width=1\textwidth]{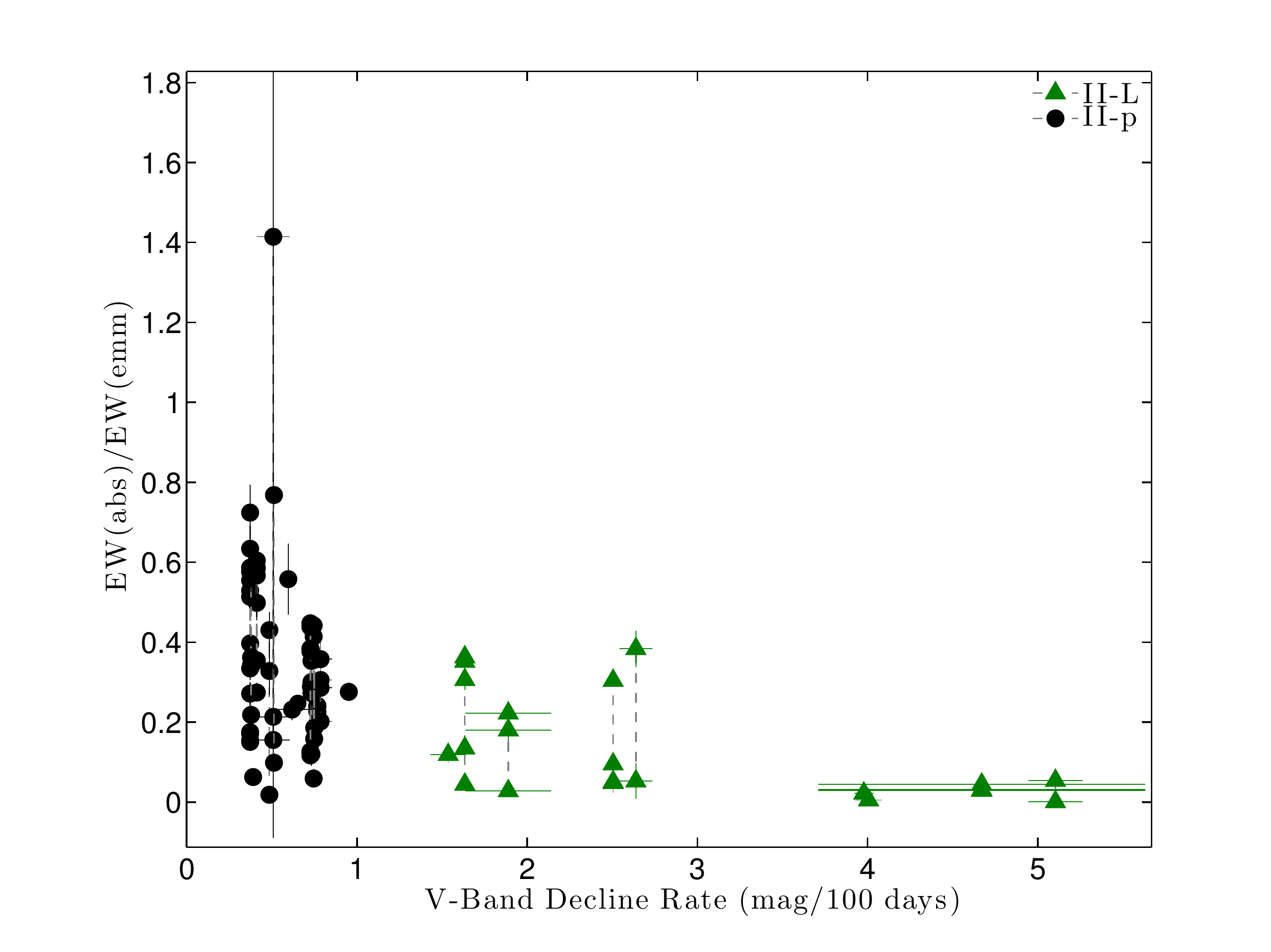}
\caption{The EW ratios of H$\alpha$ absorption to emission. Dashed lines connect multiple spectra of the same SN. These ratios are typically larger in spectra of SNe~II-P, confirming the suggestion of \citet{Schlegel:1996} and the recent results of \citet{Gutierrez:2014}.}
\label{f:ae_ratio}
\end{figure*}

\citet{Schlegel:1996} suggested that SNe~II-L have typically shallower H$\alpha$ absorption compared to emission, while SNe~II-P have deeper P-Cygni absorption profiles. In Figure \ref{f:ae_ratio} we confirm this idea, as also recently shown by \citet{Gutierrez:2014}. The data used in this figure are presented in Table \ref{t:Ha}.

\begin{table}
	\tiny
\caption{${\rm H \alpha}$ Properties}
\begin{tabular}{ccc}
\hline\hline
SN Name &
Age (days)&
EW$_{\rm H \alpha}$ (abs)/EW$_{ \rm H \alpha}$ (emm) \\
\hline
1999bg	 &41.0&0.01(0.25)	\\ 
1999co	 &36.5&0.02(0.12)	\\ 
1999em	 &11.4&0.01(0.17)	\\ 
1999em	 &13.4&0.02(0.17)	\\ 
1999em	 &14.7&0.01(0.15)	\\ 
1999em	 &15.7&0.01(0.15)	\\ 
1999em	 &25.4&0.03(0.27)	\\ 
1999em	 &28.4&0.03(0.33)	\\ 
1999em	 &30.3&0.07(0.72)	\\ 
1999em	 &34.3&0.02(0.40)	\\ 
1999em	 &38.3&0.03(0.53)	\\ 
1999em	 &41.6&0.02(0.51)	\\ 
1999em	 &44.1&0.01(0.63)	\\ 
1999em	 &48.3&0.02(0.59)	\\ 
1999em	 &51.3&0.03(0.55)	\\ 
1999em	 &53.1&0.01(0.58)	\\ 
1999gi	 &29.8&0.02(0.29)	\\ 
1999gi	 &34.8&0.02(0.38)	\\ 
1999gi	 &37.8&0.02(0.44)	\\ 
2000bs	 &74.3&0.09(0.56)	\\ 
2000dc	 &20.4&0.00(0.03)	\\ 
2000dc	 &61.4&0.03(0.18)	\\ 
2000dc	 &79.3&0.03(0.22)	\\ 
2000dj	 &25.0&0.02(0.23)	\\ 
2000dj	 &35.1&0.05(0.24)	\\ 
2000dj	 &53.0&0.02(0.24)	\\ 
2000dj	 &61.0&0.02(0.20)	\\ 
2001bq	 &55.4&0.06(0.33)	\\ 
2001bq	 &72.4&0.05(0.43)	\\ 
2001bq	 &20.2&0.00(0.02)	\\ 
2001cm	 &26.3&0.03(0.12)	\\ 
2001cm	 &43.3&0.04(0.27)	\\ 
2001cm	 &54.2&0.02(0.30)	\\ 
2001cm	 &81.1&0.02(0.35)	\\ 
2001cy	 &21.5&0.01(0.04)	\\ 
2001cy	 &32.4&0.01(0.13)	\\ 
2001cy	 &59.4&0.03(0.31)	\\ 
2001cy	 &79.2&0.03(0.36)	\\ 
2001cy	 &87.3&0.03(0.35)	\\ 
2001do	 &11.2&0.03(0.05)	\\ 
2001do	 &30.0&0.02(0.05)	\\ 
2001do	 &31.0&0.02(0.09)	\\ 
2001do	 &39.1&0.02(0.30)	\\ 
2001fa	 &30.5&0.01(0.02)	\\ 
2001x	 &26.8&0.02(0.27)	\\ 
2001x	 &34.3&0.01(0.60)	\\ 
2001x	 &35.9&0.01(0.35)	\\ 
2001x	 &52.3&0.01(0.57)	\\ 
2001x	 &66.7&0.02(0.59)	\\ 
2001x	 &82.7&0.04(0.50)	\\ 
2002ca	 &19.2&0.04(0.20)	\\ 
2002ca	 &31.6&0.05(0.36)	\\ 
2002ca	 &48.6&0.05(0.31)	\\ 
2002ca	 &89.4&0.02(0.29)	\\ 
2002gd	 &24.8&1.50(1.41)	\\ 
2002gd	 &50.0&0.01(0.16)	\\ 
2002gd	 &31.8&0.02(0.21)	\\ 
2002hh	 &44.3&0.03(0.23)	\\ 
2003hf	 &46.8&0.01(0.00)	\\ 
2003hl	 &12.4&0.01(0.06)	\\ 
2003hl	 &73.3&0.02(0.41)	\\ 
2003hl	 &68.0&0.02(0.44)	\\ 
2003iq	 &20.8&0.01(0.19)	\\ 
2003iq	 &46.8&0.01(0.36)	\\ 
2003iq	 &15.5&0.01(0.16)	\\ 
2003z	 &10.1&0.01(0.06)	\\ 
2004du	 &44.5&0.02(0.28)	\\ 
2004et	 &19.6&0.01(0.13)	\\ 
2004et	 &20.7&0.00(0.12)	\\ 
2004et	 &57.6&0.00(0.45)	\\ 
2004et	 &57.6&0.02(0.38)	\\ 
2004et	 &80.7&0.00(0.44)	\\ 
2004et	 &52.0&0.02(0.38)	\\ 
2005ay	 &29.3&0.02(0.22)	\\ 
2005ay	 &35.1&0.02(0.34)	\\ 
2005ay	 &35.1&0.02(0.34)	\\ 
2005ay	 &35.1&0.01(0.36)	\\ 
2005ay	 &63.2&0.04(0.56)	\\ 
2005cs	 &16.0&0.00(0.10)	\\ 
2005cs	 &38.0&0.02(0.77)	\\ 
2005dq	 &12.3&0.04(0.05)	\\ 
2005dq	 &73.9&0.05(0.38)	\\ 
2007fz	 &23.1&0.01(0.05)	\\ 
2007fz	 &29.9&0.01(0.00)	\\ 
2008fq	 &46.4&0.00(0.03)	\\ 
2008fq	 &57.2&0.01(0.04)	\\ 
2008fq	 &70.2&0.01(0.03)	\\ 
\hline
\label{t:Ha}
\end{tabular}
\end{table}

H$\beta$ absorption is stronger than H$\alpha$ absorption, but much weaker than in SNe~II-P. Metal lines that we identified in the spectra of SNe~II-P are also present here, and appear a bit stronger (see, e.g., O\,I $\lambda$7774) and are generally more prominent than H$\alpha$. The strong metal lines are a consequence of the high density at the photosphere expected for a rapid recombination wave. The Na\,I\,D line appears wider than its counterpart in SN~II-P spectra, suggesting either a larger velocity dispersion, or perhaps more likely, a greater contribution from the He\,I $\lambda$5876 line.

The superluminous SN\,2008es is exceptional in that it shows no features at all through day 34 after explosion. The spectrum on day 73 is more developed, but with no clear H$\alpha$ absorption. \citet{Miller:2009} also note the lack of features during early epochs, and the absence of H$\alpha$ absorption is supported by their additional spectra. They suggest that the muted amplitude of the spectral features can be explained through the ``top-lighting effect'' described by \citet{Branch:2000}. This effect rescales the amplitude of the lines by illumination from continuum emission that originates from interaction with the circumstellar medium.

\begin{figure*}
\centering
\includegraphics[width=1\textwidth]{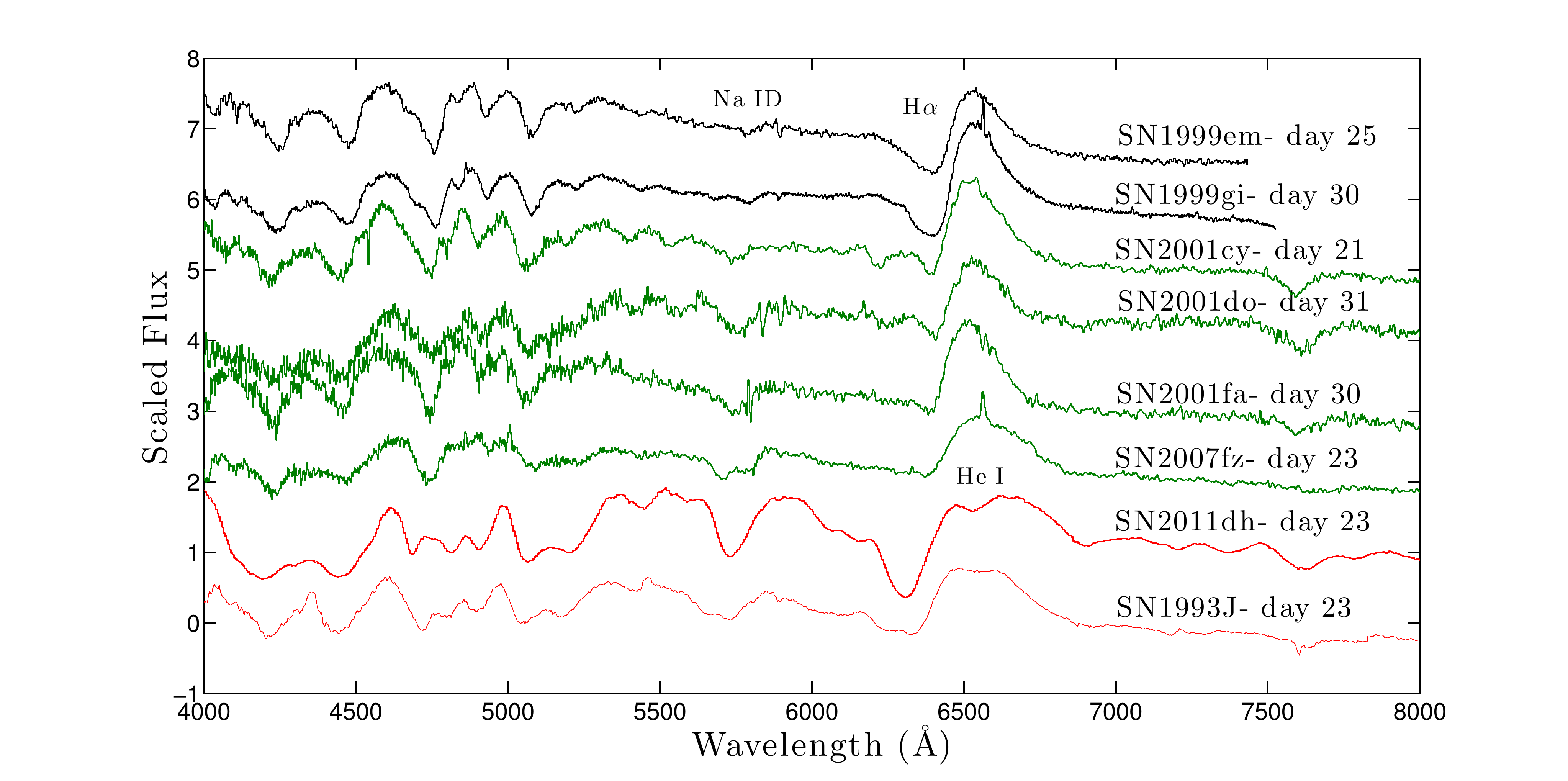}
\caption{Spectroscopic comparison between SNe\,II-P, II-L, and IIb  around days 20--30. Surprisingly, the Type IIb SN\,2011dh has a strong H$\alpha$ absorption that is equivalent to (if not stronger than) that in both SNe~II-P. The He\,I/Na\,I\,D line behaves as expected, gradually increasing in strength as we go from SNe~II-P to IIb, owing to higher helium abundance in SNe~II-L and IIb. An interesting feature is the additional absorption dip seen in the H$\alpha$ P-Cygni emission component of both SNe~IIb, which is absent in spectra of SNe~II-L and II-P. This feature can probably be associated with He\,I $\lambda$6678. Also evident is the wider P-Cygni emission in spectra of SNe~IIb than in SNe~II-P or II-L.}
\label{f:IIL_IIb}
\end{figure*}

In Section \ref{s:IIPLb} we point out the photometric resemblance of SNe\,2007fz and 2001fa to typical SNe~IIb. This raised the suspicion that they were misclassified SNe~IIb. We compare their spectra (days 23 and 30) to spectra of SNe~II-P and II-L from our joint sample at similar epochs, and to two SNe~IIb (SN\,1993J and 2011dh). Clear differences are evident among the three classes in Figure \ref{f:IIL_IIb}. Since SNe~II-L have weaker H$\alpha$ absorption compared with SNe~II-P, one could expect SNe~IIb to have even weaker hydrogen lines. Surprisingly, though, the opposite is true: SN\,2011dh has strong H$\alpha$ absorption that is equivalent to (if not stronger than) that in both SNe~II-P. The H$\alpha$ line in SN\,1993J is weaker than in SN\,2011dh, but still at least as strong as those of the SNe~II-L. The He\,I/Na\,I\,D line behaves as expected, gradually increasing in strength as we go from SNe~II-P to IIb, probably due to higher helium abundance. 

The most distinctive difference, however, is probably the additional absorption dip seen in the H$\alpha$ P-Cygni emission component, appearing in both of the SNe~IIb but in none of the SNe~II-L or II-P. This feature can probably be associated with He\,I line at 6678\,\AA. The P-Cygni emission also appears somewhat broader in both SNe~IIb (and also slightly in SN\,2007fz), pointing to a large velocity dispersion. The spectra of SNe\,2007fz and 2001fa (particularly at later phases) show more resemblance to the spectra of the other SNe\,II-L than to those of SNe\,IIb. 

However, what if SNe\,2007fz or 2003hk, for which we only have early-time spectra, later evolved to show no hydrogen, and therefore retroactively became SNe\,IIb? This might be a purely semantic question, since the diversity of progenitor systems and SN characteristic do not allow one to draw any direct conclusion from such a transition or lack thereof. More objects, with better sampling of the spectroscopic evolution, is required to answer that question and better understand how much the photometric and spectroscopic classes overlap.

\subsection{He Abundance Tracers}
\label{s:tracers}

If indeed SNe~II-L come from massive stars with thin hydrogen envelopes, and helium contributes to the different light-curve shapes, we should see clear signs of this in the spectra as well. The strongest helium line --- He\,I line at 5876\,\AA\ --- is unfortunately blended with Na\,I\,D $\lambda$5896. Other helium lines in our spectral range are too weak and usually absent. We must therefore use a proxy for a helium-rich composition. 

The two scenarios of hydrogen-rich and helium-rich envelopes are addressed in detail by \citet{Hatano:1999}, who calculated the local thermodynamic equilibrium (LTE) optical depths of the most probable ions in the Sobolev approximation as a function of temperature. The hydrogen-rich scenario describes a typical SN~II-P with a large hydrogen envelope, while the helium-rich composition represents the stripped SNe of Type Ib. A somewhat intermediate composition would be achieved in progenitors which have not lost their entire outer hydrogen envelope, but are still hydrogen deficient. In such a case, the He\,I and He\,II lines should become stronger owing to the high abundance of helium. Likewise, metal lines become more prominent as a result of the increasing density near the photosphere. From \citet{Hatano:1999} we find that O\,I is the only ion in our spectral range (apart from helium ions) that becomes more prominent when transitioning from a H-rich to a He-rich atmosphere. 

\begin{figure*}
\centering
\includegraphics[width=1\textwidth]{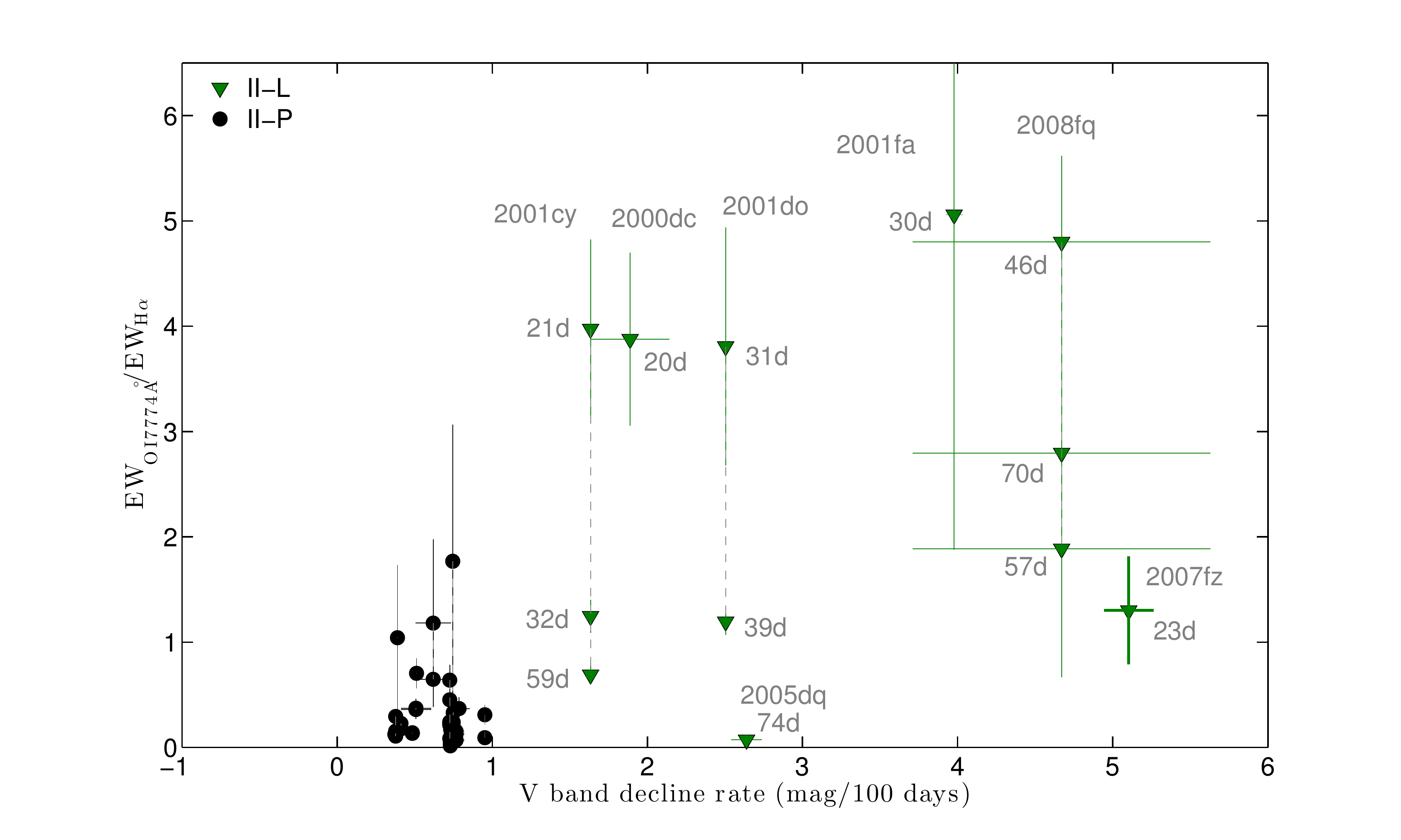}
\caption{The EW ratios of H$\alpha$ and O\,I $\lambda$7774 vs. the $V$-band decline rate. We indicate the epoch at which the EWs are measured for the SNe~II-L. Dashed lines connect multiple spectra of a same SN. One can see a correlation between decline rate and the relative strength of the O\,I $\lambda$7774 line. }
\label{f:Idecline_EW}
\end{figure*}

We measure the equivalent width (EW) of the absorption component of O\,I $\lambda$7774 for our sample, and scale it by the EW of H$\alpha$ in order to get a handle on the relative abundance. Both H$\alpha$ and O\,I $\lambda$7774 are isolated lines, between which the continuum is roughly flat and hardly varies. If our above analysis is correct, then SNe II-L (and generally objects with declining light curves) should have stronger O\,I to H$\alpha$ ratios. In Figure \ref{f:Idecline_EW} we show this ratio as a function of the $V$-band decline per 100 days. SNe~II-L typically have a higher ratio, and it weakens with time owing to the increasing strength of H$\alpha$. The data used for this figure are presented in Table \ref{t:EW_HO}.

\begin{table}
	\tiny
\caption{${\rm H \alpha}$ vs. O\,I\,7774\,\AA}
\begin{tabular}{ccc}
\hline\hline
SN Name &
Age (days)&
EW(H$\rm \alpha$)/EW(O I 7774\,\AA) \\
\hline
1999d	 &19.7&0.52(0.20)	\\ 
1999em	 &41.6&0.12(0.01)	\\ 
1999gi	 &34.8&0.04(0.01)	\\ 
1999gi	 &37.8&0.02(0.01)	\\ 
2000dc	 &20.4&3.88(0.82)	\\ 
2000dj	 &25.0&0.07(0.02)	\\ 
2000dj	 &35.1&0.13(0.07)	\\ 
2000dj	 &53.0&0.15(0.04)	\\ 
2001bq	 &55.4&0.14(0.07)	\\ 
2001bq	 &72.4&0.13(0.03)	\\ 
2001cm	 &81.1&0.10(0.02)	\\ 
2001cm	 &109.0&0.17(0.02)	\\ 
2001cy	 &21.5&3.97(0.85)	\\ 
2001cy	 &32.4&1.25(0.16)	\\ 
2001cy	 &59.4&0.69(0.09)	\\ 
2001do	 &31.0&3.81(1.13)	\\ 
2001do	 &39.1&1.19(0.13)	\\ 
2001fa	 &30.5&5.06(3.18)	\\ 
2001hg	 &35.1&0.15(0.06)	\\ 
2001hg	 &79.1&0.10(0.03)	\\ 
2001hg	 &40.5&0.15(0.04)	\\ 
2001x	 &26.8&0.23(0.03)	\\ 
2001x	 &35.9&0.18(0.02)	\\ 
2002an	 &28.1&2.73(1.35)	\\ 
2002an	 &31.2&2.18(0.34)	\\ 
2002an	 &56.0&1.06(0.20)	\\ 
2002bx	 &30.1&0.56(0.14)	\\ 
2002bx	 &30.1&0.55(0.13)	\\ 
2002bx	 &47.0&0.19(0.06)	\\ 
2002bx	 &59.0&0.19(0.04)	\\ 
2002ca	 &31.6&0.37(0.11)	\\ 
2002gd	 &50.0&0.36(0.09)	\\ 
2002gd	 &31.8&0.37(0.09)	\\ 
2002hh	 &4.3&1.18(0.80)	\\ 
2002hh	 &44.3&0.65(0.13)	\\ 
2003gd	 &85.0&0.07(0.01)	\\ 
2003gd	 &92.9&0.10(0.02)	\\ 
2003hl	 &12.4&1.77(1.30)	\\ 
2003hl	 &68.0&0.15(0.02)	\\ 
2003iq	 &20.8&0.33(0.07)	\\ 
2003iq	 &46.8&0.20(0.02)	\\ 
2003iq	 &15.5&0.24(0.07)	\\ 
2003z	 &10.1&1.04(0.69)	\\ 
2004du	 &44.5&0.31(0.09)	\\ 
2004du	 &100.3&0.09(0.02)	\\ 
2004et	 &19.6&0.64(0.14)	\\ 
2004et	 &20.7&0.45(0.06)	\\ 
2004et	 &57.6&0.23(0.01)	\\ 
2004et	 &57.6&0.21(0.02)	\\ 
2004et	 &80.7&0.08(0.01)	\\ 
2004et	 &52.0&0.24(0.02)	\\ 
2005ay	 &29.3&0.29(0.05)	\\ 
2005ay	 &35.1&0.16(0.03)	\\ 
2005ay	 &63.2&0.11(0.01)	\\ 
2005cs	 &16.0&0.70(0.14)	\\ 
2005dq	 &73.9&0.07(0.03)	\\ 
2007fz	 &23.1&1.30(0.51)	\\ 
2008fq	 &46.4&4.80(0.82)	\\ 
2008fq	 &57.2&1.89(1.22)	\\ 
2008fq	 &70.2&2.80(1.33)	\\ 
\hline
\label{t:EW_HO}
\end{tabular}
\end{table}

\section{Velocity Evolution}

We measure the H$\alpha$ and H$\beta$ velocities, along with the photospheric velocities from the minimum of the Fe\,II $\lambda$5169 line. The measurements and their uncertainties are obtained as described by F14a. We find that the photospheric velocities of the SNe~II-L are generally higher than those of the SNe~II-P throughout their evolution. However, the velocities of the SN~II-L hydrogen lines start off lower, but after $\sim 50$ days they surpass the velocities of the SNe~II-P, which seem to decline faster over time.

\begin{figure*}
\centering
\includegraphics[width=1\columnwidth]{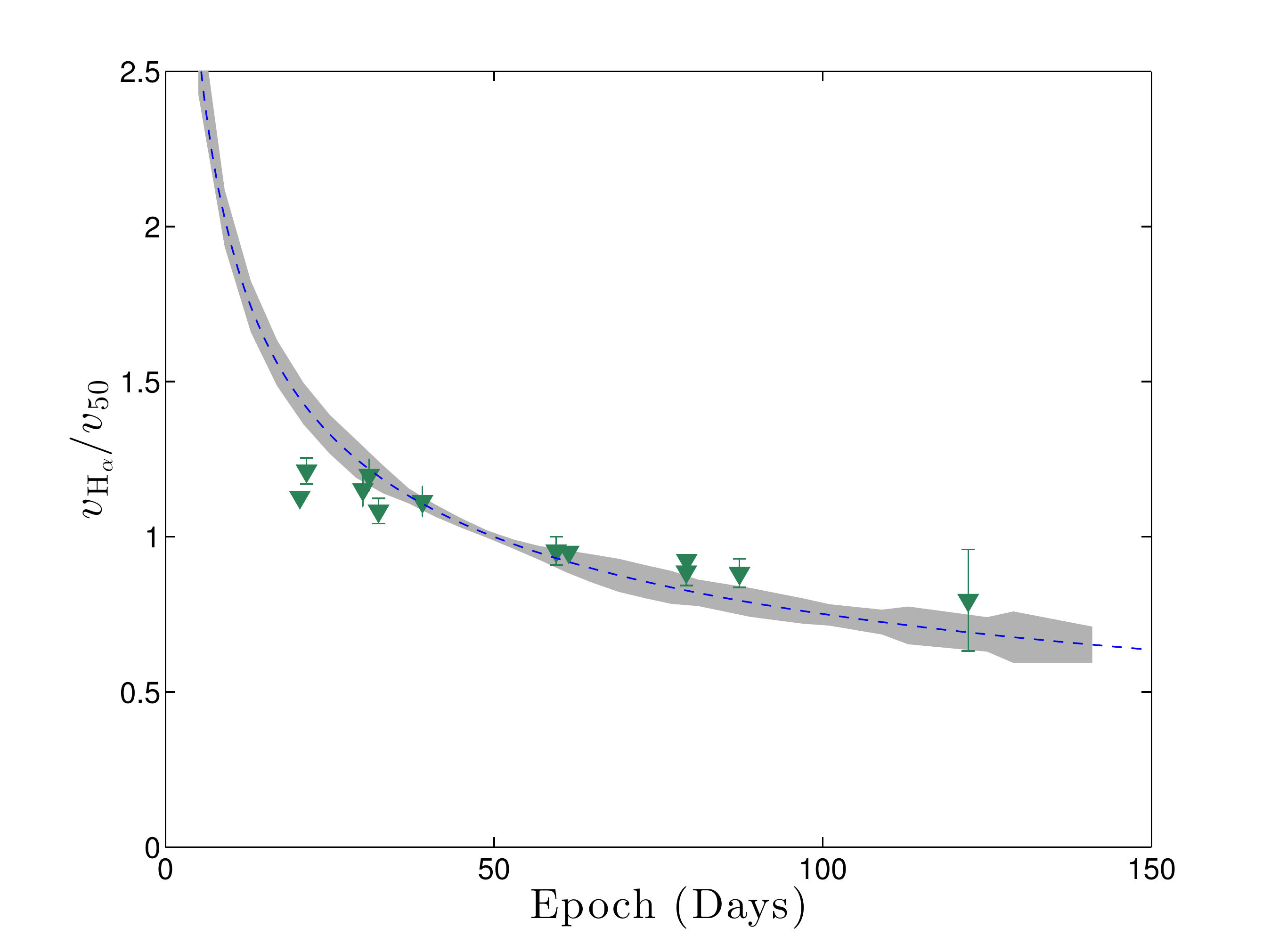}
\includegraphics[width=1\columnwidth]{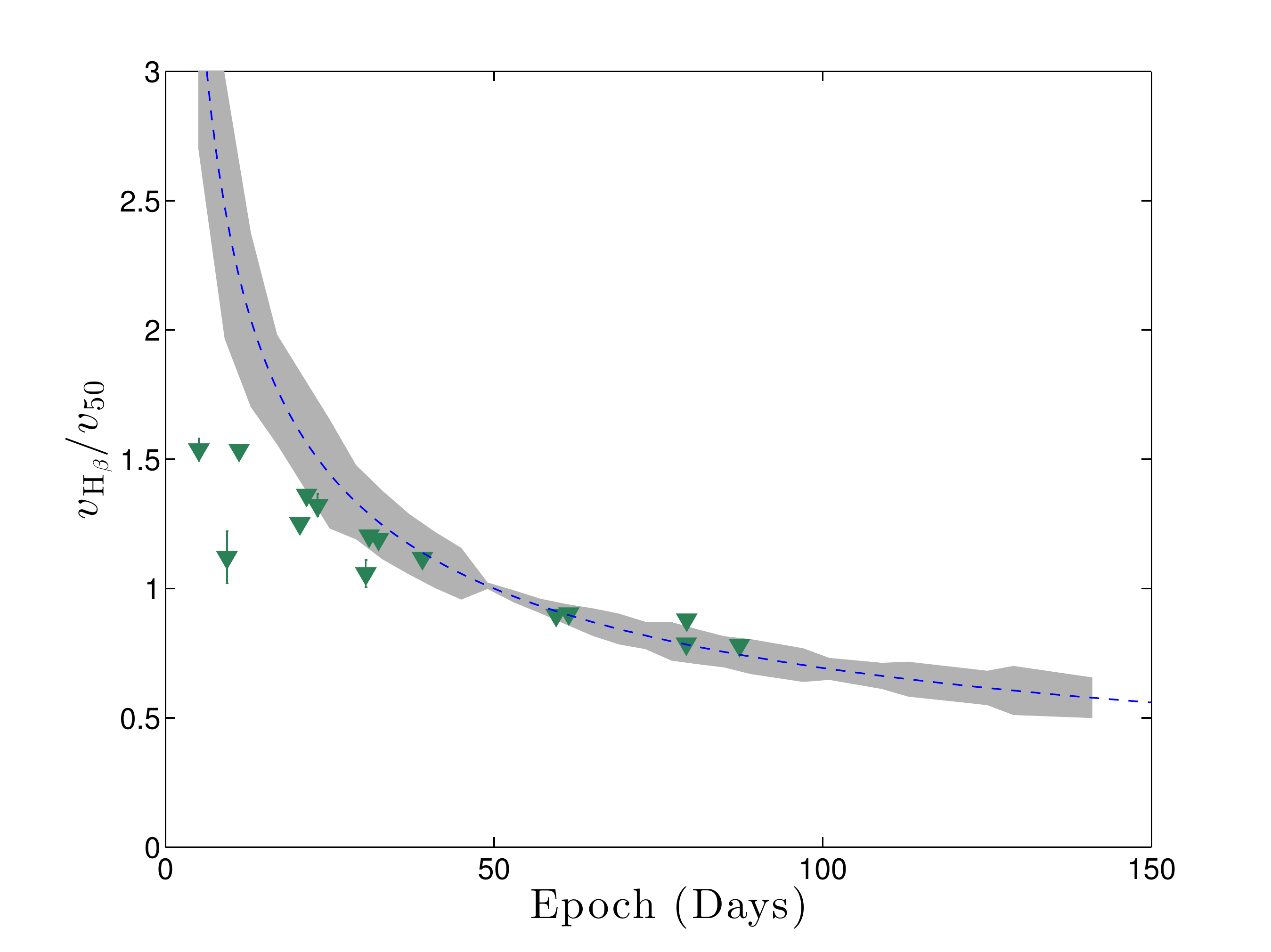}
\includegraphics[width=1\columnwidth]{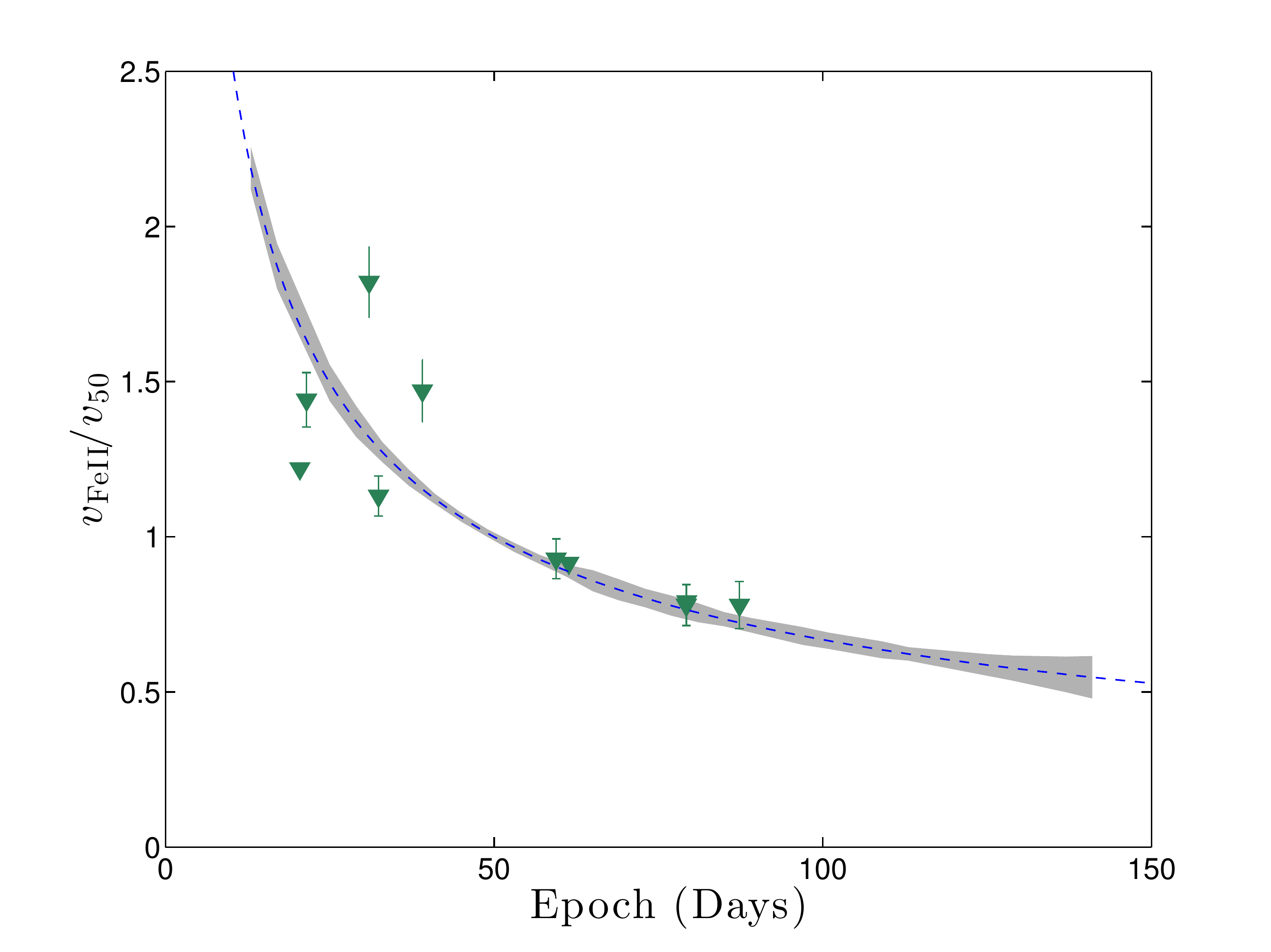}

\caption{H$\alpha$ (top-left panel), H$\beta$ (top-right), and Fe\,II $\lambda$5169 (bottom) velocities normalised by $v_{50}$, the velocity on day 50. The dashed line is the power law found by F14a for a sample of 23 SNe~II-P, and the shaded area represents the mean distance of the SN~II-P velocities from this line. One can see that the hydrogen velocities do not follow the power law. However, the Fe\,II measurements show a similar evolution with significant scatter. }\label{f:v_t}
\end{figure*}

In Figure \ref{f:v_t} we show the velocities of the three lines as a function of time, scaled by the velocity on day 50. We also plot the power laws obtained for the sample of 23 SNe~II-P by F14a. It is clear that the hydrogen velocities, especially H$\beta$, do not follow the same power law as the SNe~II-P at early epochs, and the time dependence of their velocities is quite weak. 

This can be explained as follows. As the photosphere recedes inward, it reveals hydrogen-poor layers. Therefore, the absorption comes mostly from external (and faster) layers, and not from the photosphere. As a consequence, unlike in SNe~II-P during the plateau phase, the time dependence does not result from inner and slower layers being exposed. A similar process occurs for SNe II-P at later times. The iron line, however, which (at least in SNe~II-P) might be a better tracer of the photosphere, does seem to generally follow the power law derived from SNe~II-P, albeit with significant scatter.

\section{Conclusions}

In this work we analyze a sample of 11 SNe~II-L, one of the largest such samples ever studied. We define this subclass as being those SNe~II having a $V$-band light curve that declines from peak brightness by more than 0.5\,mag in the first 50 days. We demonstrate that SNe~II-L have a spectroscopic evolution that is mostly distinct from other SN types. This however does not argue against the photometric continuum that \citet{Anderson:2014} and \citet{Sanders:2014} find between SNe~II-P and II-L. Furthermore, it seems that photometrically, SNe~II-L are closely related to SNe~IIb.

In addition to their more rapid photometric decline and higher peak luminosities (by about 1.5\,mag), we identify three spectroscopic characteristics that are common to most SNe~II-L, as follows.

\begin{itemize}
\item	SNe~II-L typically have few features and comparatively weak hydrogen absorption lines throughout most of their lifetimes, as well as a larger ratio of H$\alpha$ emission to absorption than SNe~II-P \citep{Schlegel:1996,Gutierrez:2014}. 
\item  Strong O\,I $\lambda$7774 absorption relative to H$\alpha$ at early times, compared with SNe~II-P. It seems that if caught within $\sim 3$ weeks past maximum brightness, the ratio is $> 1.0$, whereas for SNe~II-P it is $< 1.0$. Using the O\,I $\lambda$7774 line as a proxy for a helium-rich envelope, we find that the helium abundance correlates with the photometric decline rate, and that indeed SNe~II-L are more helium rich, and hence hydrogen poor. 
\item More slowly evolving H$\beta$ velocity.  In particular, SN~II-P velocities decrease from maximum light relative to day 50 by a factor of 3, whereas SN~II-L velocities generally drop by $\lesssim 50$\%. This is expected if the line is formed in the outer layers of the ejecta rather than in a thick, gradually exposed, hydrogen envelope.

\end{itemize}

All of these characteristics might suggest that SNe~II-L come from progenitors that are poorer in hydrogen than those of SNe~II-P. Along with their possible tendency to suffer from more dust extinction, this trend can be possibly explained if SNe~II-L come from stars with initial masses higher than $\sim 15$\,M$_{\odot}$.

\section*{Acknowledgments}

We thank I. Arcavi, A. Gal-Yam, E. Nakar, and D. Maoz for helpful comments on this manuscript. Some of the data presented herein were obtained at the W. M. Keck Observatory, which is operated as a scientific partnership among the California Institute of Technology, the University of California, and NASA; the Observatory was made possible by the generous financial support of the W. M. Keck Foundation. The Kast spectrograph on the Shane 3-m reflector at Lick Observatory resulted from a generous donation made by Bill and Marina Kast. We thank the dedicated staffs of the Lick and Keck Observatories for their assistance.  This research made use of the Weizmann interactive supernova data repository (\texttt{www.weizmann.ac.il/astrophysics/wiserep}), as well as the NASA/IPAC Extragalactic Database (NED) which is operated by the Jet Propulsion Laboratory, California Institute of Technology, under contract with NASA.

KAIT (at Lick Observatory) and its ongoing operation were made possible by donations from Sun Microsystems, Inc., the Hewlett-Packard Company, AutoScope Corporation, Lick Observatory, the NSF, the University of California, the Sylvia \& Jim Katzman Foundation, and the TABASGO Foundation. D.P. acknowledges support from the Alon fellowship for outstanding young researchers, and the Raymond and Beverly Sackler Chair for young scientists. D.C.L. acknowledges support from NSF grants AST-1009571 and AST-1210311. J.M.S. is supported by an NSF Astronomy and Astrophysics Postdoctoral Fellowship under award AST-1302771. A.V.F.'s group at UC Berkeley has received generous financial assistance from the Christopher R. Redlich Fund, the Richard and Rhoda Goldman Fund, the TABASGO Foundation, and the NSF (most recently through grants AST-0908886 and AST-1211916).

\bibliography{MYbib}
\bibliographystyle{mn2e}

\end{document}